\documentclass[fleqn,usenatbib]{mnras}

\usepackage{newtxtext,newtxmath}
\usepackage{multirow}
\usepackage{array}
\usepackage{mathtools}
\usepackage{comment}

\newcolumntype{R}[1]{>{\raggedleft\arraybackslash}p{#1}}

\usepackage[T1]{fontenc}

\DeclareRobustCommand{\VAN}[3]{#2}
\let\VANthebibliography\thebibliography
\def\thebibliography{\DeclareRobustCommand{\VAN}[3]{##3}\VANthebibliography}

\usepackage{graphicx}	
\usepackage{amsmath}	

\newcommand{\Eq}{Equation~}
\newcommand{\Eqs}{Equations~}
\newcommand{\Fig}{Fig.~}
\newcommand{\Figs}{Figs.~}
\newcommand{\Sec}{Section~}

\newcommand{\mstare}{{M^*}}

\newcommand{\Msun}{M_{\odot}}
\newcommand{\sigmale}{\sigma_{los}(R)}
\newcommand{\sigmal}{$\sigma_{los}(R)$}
\newcommand{\sigmape}{\sigma_{ap}(R)}
\newcommand{\sigmap}{$\sigma_{ap}(R)$}
\newcommand{\sigmazM}{$\sigma_{ap}[M^*,z]$}

\newcommand{\Sersic}{S\'{e}rsic\,}

\title[Modelling of velocity dispersion]{The weak dependence of velocity dispersion on disk fractions, mass-to-light ratio and redshift: Implications for galaxy and black hole evolution}

\author[C. Marsden et al.]{
Christopher Marsden,$^{1}$\thanks{E-mail: c.marsden@soton.ac.uk}
Francesco Shankar,$^{1}$\thanks{E-mail: F.Shankar@soton.ac.uk}, Mariangela Bernardi$^{2}$, Ravi K. Sheth$^{2}$, Hao Fu$^{1}$, \newauthor
Andrea Lapi$^{3}$ 
\\
\\
$^{1}$School of Physics and Astronomy, University of Southampton, Highfield, SO17 1BJ, UK\\
$^{2}$Department of Physics and Astronomy, University of Pennsylvania, Philadelphia, PA 19104, USA\\
$^{3}$SISSA, Via Bonomea 265, I-34136 Trieste, Italy\\
}

\date{Accepted XXX. Received YYY; in original form ZZZ}

\pubyear{2021}

\begin{document}
\label{firstpage}
\pagerange{\pageref{firstpage}--\pageref{lastpage}}
\maketitle

\begin{abstract}

Velocity dispersion ($\sigma$) is a key driver for galaxy structure and evolution. We here present a comprehensive semi-empirical approach to compute $\sigma$ via detailed Jeans modelling assuming both a constant and scale-dependent mass-to-light ratio $\mstare/L$. We compare with a large sample of local galaxies from MaNGA and find that both models can reproduce the Faber-Jackson (FJ) relation and the weak dependence of $\sigma$ on bulge-to-total ratio $B/T$ (for $B/T\gtrsim 0.25$). The dynamical-to-stellar mass ratio within $R\lesssim R_e$ can be fully accounted for by a gradient in $\mstare/L$. We then build velocity dispersion evolutionary tracks \sigmazM\ (within an aperture) along the main progenitor dark matter haloes assigning stellar masses, effective radii and \Sersic\ indices via a variety of abundance matching and empirically motivated relations. We find: 1) clear evidence for downsizing in \sigmazM\ along the progenitor tracks; 2) at fixed stellar mass $\sigma\propto(1+z)^{0.2-0.3}$ depending on the presence or not of a gradient in $\mstare/L$. We extract \sigmazM\ from the TNG50 hydrodynamic simulation and find very similar results to our models with constant $\mstare/L$. The increasing dark matter fraction within $R_e$ tends to flatten the \sigmazM\ along the progenitors at $z \gtrsim 1$ in constant $\mstare/L$ models, while \sigmazM\ have a steeper evolution in the presence of a stellar gradient. We then show that a combination of mergers and gas accretion are likely responsible for the constant or increasing \sigmazM\ with time. Finally, our \sigmazM\ are consistent with a nearly constant and steep $M_{bh}-\sigma$ relation at $z\lesssim 2$, with black hole masses derived from the $L_X-\mstare$ relation.
\end{abstract}

\begin{keywords}
Galaxies -- galaxies: evolution -- galaxies: fundamental parameters
\end{keywords}



\section{Introduction}
\label{sec:Intro}

Central velocity dispersion \sigmap, usually measured within an aperture comparable to the effective radius of the galaxy $R\lesssim R_e$, is a key property in interpreting galaxy structure and evolution. Galaxy scaling relations appear much tighter and linear when expressed in terms of \sigmap\ than, e.g., stellar mass or size \citep[e.g.,][]{Bernardi05,Bernardi11,Barone18}. 
In addition, analysis of residuals from scaling relations has revealed velocity dispersion as the most fundamental galaxy property linked to the mass of the central supermassive black hole (e.g. \citealt{Bernardi07,Shankar2016,vdB16,Shankar2017,Marsden2020}), with a weak evolution along cosmic time \citep[e.g.,][]{Shankar09,Shen15}. All of these pieces of observational evidence point to velocity dispersion as one of the main probes of both galaxy and supermassive black hole growth and assembly.

Galaxies appear to be progressively more compact at earlier epochs, with the more massive early-type galaxies showing the strongest evolution in their structural parameters \citep[e.g.,][]{Trujill06,vanDokkum08,Damja11,Cimatti12,HuertasCompany2013,Van14,Faisst17,Mowla19}. One of the most popular ways to reconcile the local and high redshift observations is to invoke galaxy-galaxy mergers, especially if minor and ``dry'' (gas-poor), which models predict to be frequent and effective enough to increase the sizes, \Sersic\ indices and to decrease velocity dispersions \citep[e.g.,][]{Robertson06,Covington08,vanDokkum08,Naab09,Nipoti03,Nipoti09,Oser10,Shankar10,Shankar10age,Convington11,Nipoti12,Oser12,Shankar13sizes,Lapi18,Zanisi21}. However, additional processes can have a significant impact on modifying the structure and dynamics of galaxies due to, e.g., feedback from a central active galactic nuclei (AGN) and/or internal star formation \citep[e.g.,][]{Fan08,Fan10,Ragone11}, or even the appearance of newly-formed larger galaxies that may drive the observed evolution in a statistical sense \citep[e.g.,][]{Carollo13,Damja15,Shankar15,Fagioli16}. Indeed, velocity dispersion is predicted to grow steadily with star formation, but in a non-linear fashion in the presence of dry (gas-poor) galaxy-galaxy mergers \citep[e.g.,][]{Robertson06,Naab2009, Nipoti09, Oser12}. Numerical simulations have shown that dry major mergers are expected to have a negligible impact on velocity dispersions, which are instead predicted to decrease due to the cumulative effect of minor dry mergers. Stochastic evolutionary sequences driven by strictly dry mergers are thus expected, on average, to steadily decrease velocity dispersions in galaxies with cosmic time. In merger-free galaxy growth histories dominated by passive or weak evolution, velocity dispersions should instead retain memory of the formation epochs of the galaxies \citep[e.g.,][]{Granato2004,Cook10,Lapi18}. In this framework, using an analytic model \citet{Ciras05} were able to reproduce the full local velocity dispersion function of early-type galaxies by linking velocity dispersion to the circular velocity of the halo \citep[e.g.,][]{Ferrarese2002} at the epoch of the main burst of star formation. Dissecting the $\sigma$ evolutionary tracks along the progenitors and/or at fixed stellar mass, which is the main aim of this work, can thus shed light on the different channels controlling galaxy assembly.

In addition, despite a difference in size of several orders of magnitude, in all galaxies observed with high enough sensitivity, the mass of the central supermassive black holes appears tightly correlated with the host galaxy stellar velocity dispersion \citep[e.g.,][]{Ferrarese2000}, stellar bulge mass \citep[e.g.,][]{Marconi2003}, or even the host dark matter halo \citep{Ferrarese2002} and other properties \citep[see, e.g.,][for comprehensive reviews]{KormendyHo2013,Graham2015}. Several works \citep[e.g.,][]{Bernardi07,Hopkins07,Shankar2016, Shankar2017, Marsden2020,Feoli} showed via detailed analysis of the residuals in the supermassive black hole--galaxy scaling relations, that velocity dispersion plays a vital role \citep{Bosch_2016}, being one of the most fundamental galaxy properties related to supermassive black hole mass. These findings favour Active Galactic Nuclei (AGN) feedback models which predict a steep scaling with velocity dispersion \citep[e.g.,][]{SilkRees,Granato2004,Fabian12,King2015}, suggesting that enough energy/momentum can be transferred to the galaxy to self-regulate supermassive black hole growth, and possibly contribute to the overall structural and dynamical evolution of the host galaxy \citep[e.g. ][]{Fan2008, Fan10, Ishibashi2013, Lapi2011, Ragone2011}. However, it remains to be seen whether a sequence of stochastic galaxy mergers \citep[e.g.,][]{Hopkins2010mergers} can reproduce a tight and steep $M_{bh}-\sigma$ relation \citep[e.g.,][]{Malbon2007,Jahnke2011, VolonteriNatarajan,Hirschmann10,RicarteNat2018}. Ultimately, understanding the relative contributions of mergers, star formation and AGN activity in the evolution of supermassive black holes and their host galaxies remains a vital open question, and velocity dispersion may be one of the key elements to solve this puzzle.

Velocity dispersion is also a tracer of the distribution of dark matter in the inner regions of the galaxies, via the dynamical mass $M_{dyn}(<R)\propto R\,\sigmape^2$, which traces the gravitational influence of the total mass within $R$ on test (stellar or gas) particles. The ratio between dynamical and stellar mass within the effective radius, $M_{dyn}(<R_e)/\mstare(<R_e)$, appears to be increasing with stellar mass, $M_{dyn}\propto \mstare^{1+\alpha}$, with $\alpha \sim 0.2-0.3$ \citep[e.g.,][]{Pahre98,Padma04,Gallazzi06,Hopkins09}, which is related to the overall ``tilt'' of the fundamental plane (FP) of galaxies \citep[e.g.,][]{Djorgovski1987FP, Dressler1987FP}, where the FP terminology is usually mostly applied to earlier type galaxies \citep[e.g.,][but see also \citealt{Ferrero21}]{Bernardi2020FP}. Different effects could contribute to the tilt of the FP and more specifically to the slope of the $M_{dyn}/\mstare$ ratio, from an increasing contribution of dark matter in larger/more massive galaxies, to the effect of stellar non-homology, radial anisotropy, and/or systematic variations of the stellar population  \citep[e.g.,][]{Ciotti96,Trujillo04,Bertin06,Cappellari06,HydeBernardi09,Hopkins09,Cappellari13,
Onofrio13,Zahid16,Zahid17,Bernardi2020FP,Mould20,Eugenio21}. 

In recent years it has become evident that the stellar initial mass function (IMF) may not be a universal constant.  Systematic changes in the IMF, either across the population, or within a galaxy itself, can lead to changes in the stellar mass to light ratio.  A large body of relatively recent literature argues that the systematic increase of $M_{dyn}/\mstare$ with $\mstare$ is almost entirely due to an IMF-driven $\mstare/L$ variation across the galaxy population \citep[e.g.,][]{Cappellari06,Conroy12,Cappellari13,Lyube16,Tang17,Cappellari2017,Cappellari2018}.  These studies conclude that the IMF correlates with $\sigma_{ap}$, becoming more bottom-heavy, Salpeter-like \citep{SalpeterIMF} in galaxies with larger $\sigma_{ap}(<R_e)$, so that stellar mass estimates based on stellar population modeling which assume a universal IMF are systematically incorrect.  However, these analyses assume that the IMF, and hence $\mstare/L$ within a galaxy, is constant (In reality, even at fixed IMF, age and metallicity gradients within a galaxy will result in some $\mstare/L$ gradients.  However, for massive galaxies, these are known to be sufficiently small that they do not significantly affect Jeans-based $M_{dyn}$ estimates).  Recent work has shown that the IMF may, in fact, vary within a galaxy, becoming more bottom-heavy closer to the galaxy centre \citep[e.g.,][]{Navarro15,LaBarbera16,Lyube16,vanDokkum17,Parikh18}.  
Such IMF-driven gradients lead to more significant $\mstare/L$ gradients. This led \cite{Bernardi2018} to argue that IMF-driven $\mstare/L$ gradients within a galaxy could account for much of the perceived variation in $M_{dyn}/\mstare$, and subsequent analyses have shown that realistic gradients may indeed reconcile the $M_{dyn}$ and $\mstare$ estimates \citep[][]{Chae18,Chae19,Dominguez19,Bernardi19}. In the first part of the present work we will show that mass and dynamical modelling which include an IMF-driven $\mstare/L$ gradient (in addition to age, metallicity, etc. gradients) can indeed reproduce the full systematic increase of $M_{dyn}/\mstare$ with $\mstare$ as measured by large local galaxy samples. However, we will also show that ignoring such IMF-related effects when estimating $\mstare$, still matches some of the main $\sigma$-related observables, which is a relevant result given that estimating IMF-related effects is costly (high S/N spectra are required).  

\begin{figure*}
    \centering
    \includegraphics[width=\textwidth]{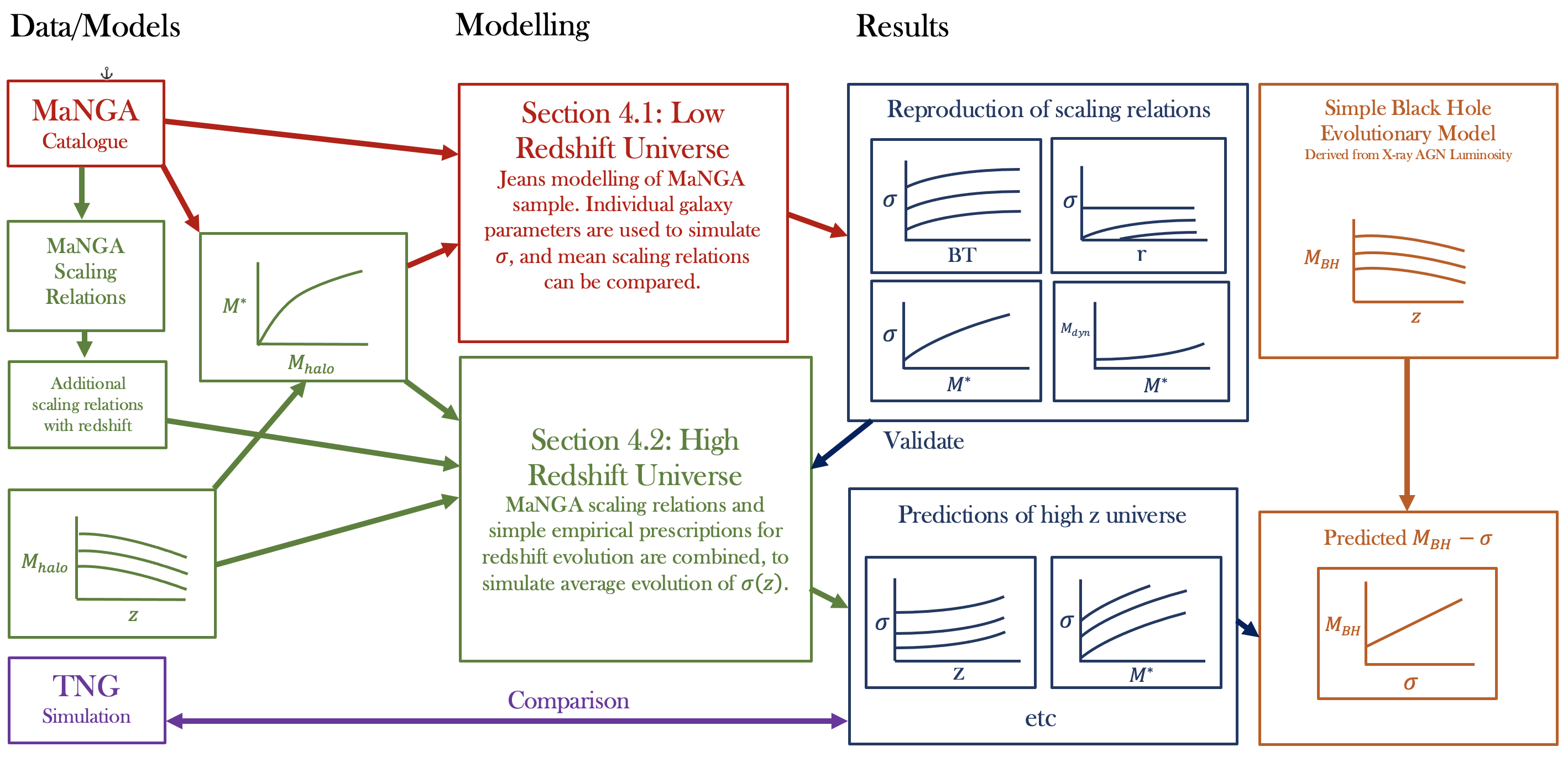}
    \caption{A visual summary of this paper. The MaNGA catalog and associated parameters are used to as the basis of our modelling of the low redshift universe. This modelling validates our methodology, which is then applied to the high redshift universe. These results are compared with the TNG simulation, and are combined with a simple black hole evolutionary model to predict the $M_{\rm BH}\!-\!\sigma$ relation and it's evolution.}
    \label{fig:structure}
\end{figure*}

More generally, in this paper we present a flexible yet accurate Jeans modelling approach developed under a variety of assumptions on: 
 the stellar profile, including wherever relevant a disk component; the inner dark matter profile; and 
 the evolution with redshift of the galactic structural properties relevant to velocity dispersion. 
We show that, whatever the exact choice of input parameters, all our models are able to reproduce the local data and make specific predictions on the relative roles of dark matter and stellar IMF, and on the evolution with redshift of the \sigmap\ which are difficult to reconcile with models based solely on dry mergers. 

The paper is structured as follows. In \Sec\ref{Data} we present our reference sample and hydrodynamic simulations. We then detail our semi-empirical modelling of velocity dispersions at both low and high redshift in \Sec\ref{Method}. We present our results in \Sec\ref{Results}, divided into low- and high-redshift Universe, discuss the consequences of our findings in light of merger models and black hole assembly in \Sec\ref{sec:Discussion}, and conclude in \Sec\ref{sec:Conclu}. Wherever required, we adopt as a reference the Planck Cosmology \citep{Planck18}, noting that none of our core results depend on the exact choice of cosmological parameters.

\section{Observational and numerical comparison data} 
\label{Data}

In what follows, we will compare our numerical models for galaxy velocity dispersions with both simulated and observed samples. 
The former comes from the TNG50 hydrodynamical simulation \citep{TNGSup2,TNGSup1,TNGMain} developed in a 50 Mpc box, resolved down to a baryonic mass of $8 \times 10^4\, \Msun$. As detailed below, from TNG50 we select a variety of central (i.e non-satellite) galaxies of different stellar masses at $z=0$, follow their evolution along their main progenitor branches, and at each timestep compute their (stellar) velocity dispersions (within the specified aperture) directly from the available stellar particle data, as described in Appendix 2.

The latter consists of a large sample of galaxies ($\sim 4700$) published by \cite{Fisher19} which was part of an early release (SDSS-DR15, \citealt{Aguado2019}) of the MaNGA (Mapping Nearby Galaxies at Apache Point Observatory; \citealt{Bundy2015}) Survey. The final release of MaNGA will provide integral field unit spectroscopy and accurate kinematic maps (stellar rotation velocity and velocity dispersion) for $\sim 10^4$ nearby ($0.03<z<0.15$) galaxies distributed across 4000~deg$^2$ and uniform over the mass range $M_* \ge 10^9-10^{12}\, M_\odot$ with no size, inclination, morphology or environmental cuts. The spectra (also called spaxels) have a wavelength coverage of $3500-10^4$\AA. Individual spaxels in the central region of a MaNGA galaxy have S/N $\sim 100$, but the vast majority of the spaxels have S/N $< 50$. Two-thirds of the galaxies have spatial coverage, at about 1~kpc resolution, to 1.5 times the half-light radius of a galaxy; the other third of the sample will have coverage to 2.5$R_e$. Early-type galaxies make up about forty percent of the sample, and will contribute about $10^6$ spaxels; spirals, the other sixty percent, contribute about fifty percent more spaxels. Velocity dispersion in MaNGA provides velocity dispersions with and without correction for the rotational component. In what follows we always use as reference ``full'' velocity dispersions, i.e., uncorrected for the rotational component, though we note that excluding the rotational component has a negligible impact on our results. 

We use the photometric parameters and bulge-to-total (B/T) decompositions published by \citet{Fisher19} which are based on single component S\'{e}rsic \citep[e.g.,][]{Sersic1968} or two-component S\'{e}rsic+Exponential light profiles. We also use morphological properties derived from supervised Deep Learning algorithms based on Convolutional Neural Networks \citep{Dominguez18} also from \citet{Fisher19}. We restrict the observational sample to only those galaxies with a reliable fit to a single \Sersic profile (\textit{flag\_fit=1}) or to a \Sersic+Exponential profile (\textit{flag\_fit=0 and 2}). We will sometimes refer to the two components as the bulge and disk components, even if there is no hint of a disk in the imaging.  For each object, we consider two definitions of stellar mass:
 one is based on the total stellar mass-to-light ratio from \citet{Mendel14}, which assumes a Chabrier \citep{ChabrierIMF} stellar initial mass function (IMF) for all galaxies;
 the other allows for a gradient in $\mstare/L$ which is driven by an IMF gradient in each object, as described in \Sec\ref{Method}. Since a galaxy with a fixed IMF may still have a gradient in $\mstare/L$ (due, e.g., to age and metallicity gradients in its stellar population) we will sometimes refer to the former as `IMF fixed' models, and the latter as `IMF gradient' models.  However, we follow common practice in ignoring the $\mstare/L$ gradients in our `IMF fixed' models, so that $\mstare/L$ in these models is constant within a galaxy, but this constant can  differ from one galaxy to another.

\section{Method} 
\label{Method}

The primary aim of the current work is to provide a robust analytic methodology to infer the average velocity dispersion across galaxy stellar mass and time \sigmazM, a key prediction for galaxy evolutionary models in a cosmological context. Our results can then be used in cosmological semi-analytic, semi-empirical models and also in lower resolution cosmological simulations, and also adopted by observational teams to calibrate velocity dispersions of galaxy samples at different epochs. To achieve this goal, we will proceed in two steps, which are visually summarised in \Fig\ref{fig:structure}. We will first start in the local Universe (upper red box in \Fig\ref{fig:structure}) from the mass-complete sample of MaNGA galaxies introduced in \Sec\ref{Data}, with measured stellar light profiles and velocity dispersions. We will use the light profile fitting results in terms of effective radius, bulge-to-disc decomposition (when adopting \Sersic-Exponential light profile fits) and \Sersic index, as inputs in our mock sample to predict, via detailed Jeans modelling, the velocity dispersion profiles, which we will compare with those actually measured in the MaNGA sample. We will make use of abundance matching relations, i.e., the (inverse) of a stellar mass-halo mass (SMHM) relation \citep[e.g.,][]{Shankar17gamma}, to assign dark matter haloes to each MaNGA galaxy. Stellar masses in the MaNGA galaxies, as discussed in \Sec\ref{Data}, are inferred either assuming a fixed IMF, or by including a stellar gradients, and we will carefully explore the impact of the latter on our results. From the predicted velocity dispersions we can then infer different relevant galaxy scaling relations such as the $M_{dyn}/\mstare$ ratio, which is informative to shed light on the FP of galaxies (\Sec\ref{sec:Intro}), or the $\sigma-\mstare$ relation (top, right panels of \Fig\ref{fig:structure}). After having identified a robust methodology to successfully reproduce the data at $z=0$, in the second part of the work we move to predicting \sigmazM\ at higher redshifts, with a focus in probing the velocity dispersion evolution of strictly \emph{central} galaxies along the progenitor galaxies and at fixed stellar mass (lower middle box in \Fig\ref{fig:structure}; details in \Sec\ref{Results-High}). To calculate calculate \sigmazM\ along the progenitors, we rely on the mean host halo mass evolutionary tracks as inferred from extensive analytic and numerical models, and assign stellar masses to haloes via abundance matching techniques. We then assign to our mock galaxies empirically informed effective radii and \Sersic indices as calibrated against a variety of observational data sets. We also explore the impact on our results of adopting effective radii derived from halo virial radii. We will then determine the evolution with redshift of the main observables calibrated at $z=0$, namely the $M_{dyn}/\mstare$ ratio and the $\sigma-\mstare$ relation (bottom, right panels of \Fig\ref{fig:structure}). We will then compare our model predictions with the outputs of cosmological hydrodynamical simulations (bottom of \Fig\ref{fig:structure}), and we will discuss, in \Sec\ref{sec:Discussion}, the putative roles of gas accretion versus dry mergers in driving the evolution in \sigmazM. We will conclude our work by predicting the evolution of the black hole mass-velocity dispersion relation, which we will obtain by coupling our \sigmazM\ evolutionary tracks with the evolution of black hole mass derived from data-driven accretion models \citep[e.g.,][]{Yang19,Shankar20}. We devote the remainder of \Sec\ref{Method} to a detailed breakdown of the steps that we follow to compute velocity dispersions at both low and high redshifts.

\subsection{Simulating Velocity Dispersion, $\sigma$.} \label{sigma}

\begin{figure*}
	\includegraphics[width=\textwidth]{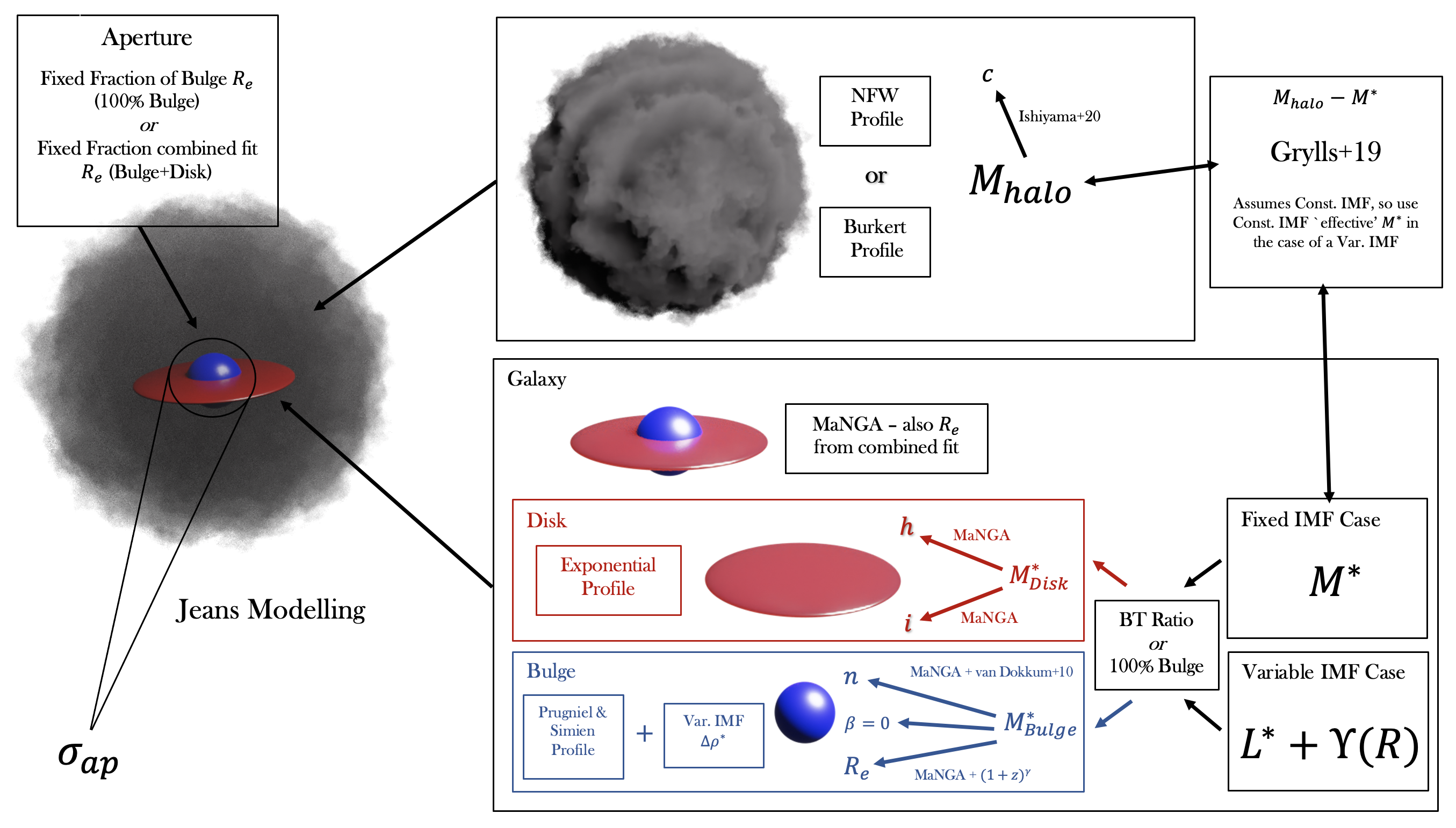}
    \caption{Cartoon image showing the contribution of the various elements of our model. Stellar masses are assigned to haloes (or vice-versa) using the relation from \citet{Grylls2019STEEL}, refined by \citet{Zanisi21}, and associated disk and bulge properties are assigned using appropriate scaling relations (see text for details). The bulge, disk and dark matter components are brought together at fixed aperture size to simulate \sigmap. This image was created using the Blender software, using assets from the Astera project \citep{Marsden2020Astera}.}
    \label{fig:Diagram}
\end{figure*}

We here detail our step-by-step methodology for computing the galaxy velocity dispersion \sigmap\, which we visually describe in \Fig\ref{fig:Diagram}. Our aim is to compute the \sigmap\ of the system galaxy+host dark matter halo within a given aperture $R$ (left side of \Fig\ref{fig:Diagram}). For the dark matter component (top panel of \Fig\ref{fig:Diagram}) we define a mass profile and a concentration extracted from N-body simulations. In the specific, we adopt as a reference the Navarro, Frenk and White (NFW) profile \citep{NFW1996} for the dark matter component $M_{halo}(r)$, with a concentration-halo mass relation from \cite{Ishiyama2020}, and we will also show the impact of switching to a cored dark matter profile \citep[][]{Burkert95}.  
For the stellar component instead (bottom panel of \Fig\ref{fig:Diagram}), for galaxies that are well described by a single-\Sersic\ light profile, we treat the stellar mass as a single dynamical component (i.e., $B/T\!=\!1$, neglecting the disk component completely), whereas for two-component \Sersic+Exp models we separately calculate the disk and bulge/spheroid velocity dispersion (within a given aperture) and then sum them in quadrature:
\begin{align} \label{combined}
    L_{ap,tot}\sigma_{ap,\,tot}^2 &= L_{ap,bulge}\sigma_{ap,\,bulge}^2 + L_{ap,disk}\sigma_{ap,\,disk}^2\ 
\end{align}
where 
$L_{ap}\equiv 2\pi \int_0^{R_{ap}} dR\,R\,I(R)$ and $L_{ap,tot}\equiv L_{ap,bulge} + L_{ap,disk}$.  
Since the bulge typically dominates on relatively smaller scales than the stellar disk, we expect our luminosity-weighted total velocity dispersion given in \Eq\ref{combined} to provide a smooth interpolation between the bulge and disk values.
Our method for computing \sigmap\ closely follows \citet{Desmond17} for models with a constant $\mstare/L$, and \citet{Bernardi2018} for models with an IMF-driven scale-dependent $\mstare/L$.  As mentioned above and as visualised in the bottom, right panels of \Fig\ref{fig:Diagram}, all the input galaxy properties such as disk and bulge radii, inclination, or \Sersic\ profile properties, are all measured quantities in the MaNGA reference sample.  In principle, the dark matter profile responds to the formation of stars in the central regions.  Recent work \citep{Paran21} shows how to include this effect, but for simplicity and consistency with previous work, we ignore it.  Moreover, if IMF-gradients can be ignored, then strong lensing measurements at different redshifts and stellar masses \citep[e.g.,][]{Sonne15} suggest that such adiabatic contraction or expansion effects are weak \citep[e.g.,][]{Dutton14,Shankar17gamma,ShankarGammaz}. Nevertheless, we will show below that all our models, irrespective of the chosen light profile and/or $\mstare/L$, predict a only weak dependence on $B/T$, in excellent agreement with what is measured in the reference observational sample. In the following subsections we discuss how we model the contribution for both the bulge and the disk.   

\subsection{Spheroids/Bulges}

\subsubsection{Solving the Jeans Equations}

We compute the velocity dispersion of a spheroid by solving the Jeans equation \citep[e.g.,][]{Jeans1915,Binney2ndEd} assuming spherical symmetry and no streaming motions:
\begin{equation}
    \label{Jeanseq}
  \frac{d[\rho(r)\sigma_r^2]}{dr} + 2\frac{\beta(r)}{r}\rho(r)\sigma_r^2 = -\rho(r)\frac{GM(r)}{r^2} \, .
\end{equation}
In \Eq\ref{Jeanseq}, $r$ is the radial distance, $\rho(r)$ is the radial (mass) density of the galaxy, $M(r)$ is the total cumulative mass within $r$, $\sigma_r$ is the radial component of the velocity dispersion, and $\beta(r)$ the radial velocity anisotropy:
\begin{equation}
\beta(r) = 1 - \frac{\sigma_t^2}{\sigma_r^2} \, ,
\end{equation}
with $\sigma_t = \sigma_\theta = \sigma_\phi$ the tangential component of $\sigma$. Assuming $\beta = 0$ corresponds to perfect isotropy, $\beta=1$ to fully radial anisotropy and $\beta\rightarrow-\infty$ fully tangential anisotropy. Unless otherwise noted, we will assume $\beta=0$ throughout this work, which is a good approximation to reproduce the MaNGA data at $z=0$. There is some precident for this, as isotropic orbits were found in the work of \cite{Merritt1987, LokasMamon2003, Katgert2004} (for a more detailed discussion of this topic, we refer the reader to \cite{MamonLokas2005}, section 3.2). We verified that none of our main results, especially the ones referring to the time evolution of velocity dispersion, change qualitatively when assuming, for example, a non-zero anisotropy $\beta \sim 0.3-0.4$ as often adopted in massive early-type galaxies \citep[e.g.,][]{MamonLokas2005,Chae19}. 
Note that if the two components in a SerExp fit are spheres with different dynamics (but the same $\beta$), then the result of writing equation~(\ref{Jeanseq}) for each of the two components and summing the two expressions is again equation~(\ref{Jeanseq}) with $\rho_{tot}\sigma_{r,tot}^2 = \rho_{Ser}\sigma_{r,Ser}^2 + \rho_{Exp}\sigma_{r,Exp}^2$.  We discuss how we treat the possibility that the Exponential component is a rotationally supported disk shortly. 

The general solution to \Eq\ref{Jeanseq} can be written as
\begin{equation}
    \label{eq:GeneralSoln}
    \rho(r)\sigma_r^2(r) = \frac{1}{f(r)}\int_{r}^{\infty} f(s) \rho(s) \frac{GM(s)}{s^2} ds ,
\end{equation}
where $f$ is the solution to the differential equation \citep[see, e.g.,][]{vanderMarel1994}
\begin{equation}
    \label{GeneralSolnComponent}
    \frac{d\ln f}{d\ln r} = 2\beta(r)\, .
\end{equation}
Projecting the velocity ellipsoid along the line of sight, it can be shown \citep{Binney1982} that the line-of-sight velocity dispersion \sigmal\ at projected radius $R$ is
\begin{equation}
    \label{eq:losFULL}
    \frac{1}{2}I(R)\sigma_{los}^2(R) = \int_R^\infty \frac{\ell(r)\sigma_r^2rdr}{\sqrt{r^2-R^2}} - R^2 \int_R^\infty \frac{\beta\ell(r)\sigma^2_rdr}{r\sqrt{r^2 - R^2}} \, .
\end{equation}
Inserting \Eq\ref{eq:GeneralSoln} and \Eq\ref{GeneralSolnComponent} into \Eq\ref{eq:losFULL}, implies
\begin{equation}
    \label{Eqsplit}
    \begin{split}
    \frac{I(R)\sigma_{los}^2(R)}{G} = 2\int_R^\infty \frac{f\ell M}{s^2}ds \int_R^s \frac{1}{f} \frac{rdr}{\sqrt{r^2 - R^2}} \\- R^2 \int_R^\infty \frac{f\ell M}{s^2}ds\int_R^s\frac{df/dr}{f^2}\frac{dr}{\sqrt{r^2-R^2}} \, .
    \end{split}
\end{equation}
On the assumption of constant anisotropy ($f(r)=r^{2\beta}$), it can be shown \citep[e.g.][Appendix A]{MamonLokas2005} that \Eq\ref{Eqsplit} reduces to
\begin{equation}
    \label{eq:sigmaLOS}
    \frac{I(R)\sigma_{los}^2(R)}{G} = 2\int_R^\infty K\left(\frac{r}{R}\right) \ell(r) M(r) \frac{dr}{r} \, ,
\end{equation}
where
\begin{equation}
\label{Kernel}
\begin{split}
    K(u) \equiv \frac{1}{2} u^{2\beta-1} \left[\left(\frac{3}{2}-\beta\right)\sqrt{\pi}\frac{\Gamma(\beta-\frac{1}{2})}{\Gamma(\beta)} \right. \\ \left.
    + \beta B\left(\frac{1}{u^2}, \beta+\frac{1}{2}, \frac{1}{2} \right) - B\left( \frac{1}{u^2}, \beta-\frac{1}{2}, \frac{1}{2}\right) \right] \, .
\end{split}
\end{equation}
In \Eq\ref{Kernel}, $B$ is the incomplete beta function in the format $B(z, a, b)$\footnote{In the third term of \Eq\ref{Kernel}, where necessary we recursively use the transformation $B(z, a, b)a = z^a (1-z)^b + (a+b)B(z, a+1, b)$.}. Various alternative expressions for $K(u)$ can be adopted, as summarised by \citet{MamonLokas2005}. Finally, the aperture velocity dispersion is computed as \citep[e.g.,][]{Cappellari06,Chae14}
\begin{equation}
    \sigma_{ap} \equiv \sigma(R_{ap}) = \sqrt{\frac{\int_0^{R_{ap}}I(r)\sigma_{los}^2rdr}{\int_0^{R_{ap}}I(r)rdr}\, }.
\end{equation}
We adopt \Eqs\ref{eq:sigmaLOS} in the case of a constant IMF and mass-to-light ratio $\Upsilon_0=\mstare/L$, which implies a linear and constant scaling between light and stellar mass and reduces computational cost as it does not require calculation of the radial velocity dispersion. In the case of a scale-dependent IMF and mass-to-light ratio $\Upsilon(R)$ instead (\Sec\ref{subsec:IMFr}), we first numerically solve \Eq\ref{eq:GeneralSoln} to derive $\sigma(r)$ and then ``light-weight'' it via \Eq\ref{eq:losFULL} to obtain the line-of-sight velocity dispersion \sigmal. 

\subsubsection{Spheroids: Mass and Stellar Profiles}
\label{subsec:profiles}

Following the formalism presented in the previous section, computing $\sigma$ requires knowledge of the projected light density profile $I(R)$, of the total cumulative mass profile $M(r)$, the 3D light density profile $\ell(r)$, and the 3D stellar density profile $\rho(r)$ which, in the case of a constant $\mstare/L$, is simply given by $\rho(r)=\Upsilon_0\ell(r)$. For a bulge or a spheroid, the projected light density profile $I(R)$ is well approximated by a \Sersic profile \citep{Sersic1968}
\begin{equation}
    I(R) = I_e \exp{\left\{-b_n \left[\left(\frac{R}{R_e}\right)^{1/n} -1\right]\right\}} \, ,
    \label{eq:sersic}
\end{equation}
where $n$ is the \Sersic\ index and $b_n$ is chosen so that the luminosity within $R_e$ is half the total luminosity. \cite{Ciotti1999} approximate $b_n$ as
\begin{equation}
    b_n \simeq 2n - \frac{1}{3} + \frac{0.09876}{n} \, .
\end{equation}
The corresponding ``deprojected'' 3D light profile $\ell(r)$ is well approximated by a similar expression \citep{Prugniel997}
\begin{equation}
\label{deprojected}
    \ell(r) = \ell_0 \left(\frac{r}{R_e}\right)^{-p_n}\exp{\left\{ -b_n\left( \frac{r}{R_e} \right)^{1/n} \right\}} \, ,
\end{equation}
where
\begin{equation}
    \ell_0 = \frac{I_0 b_n^{n(1-p_n)}}{2\pi R_e^3}\frac{\Gamma(2n)}{\Gamma[n(3-p_n)]}
\end{equation}
and
\begin{equation}
    p_n = 1 - \frac{0.6097}{n} + \frac{0.00563}{n^2}\, .
\end{equation}
When adopting a constant stellar mass to light ratio $\Upsilon_0$, both the projected and 3D stellar density profiles will be identical to the light profiles simply scaled by $\Upsilon_0$. The stellar cumulative profile $M^*(r)$ also has an analytic expression in this case which reads as
\begin{equation}
\mstare(r)=\mstare\frac{\gamma_l[n(3-p_n),b_n(r/R_e)^{1/n}]}{\Gamma[n(3-p_n)]} \, ,
    \label{eq:MSersicR}
\end{equation}
with $\gamma_l$ the (lower) incomplete gamma function. 
In the case of an IMF-driven $\mstare/L$ gradient, we instead obtain $\mstare(r)$ via direct integration of the 3D stellar mass density described in \Sec\ref{subsec:IMFr}. 

Finally, the total galaxy cumulative mass profile $M(r)$ is obtained by linear addition of the different components of stars, dark matter, gas and central black hole
\begin{equation}
M(r) = \mstare(r) + M_{halo}(r) + M_{gas}(r) + M_{bh}\, .
\label{eq:massbudget}
\end{equation}
In our reference model we will assume a single \Sersic light profile, and the corresponding $\mstare(r)$ refers to the total stellar mass of the galaxy. In models in which we explicitly distinguish the bulge and disk components of the galaxy, we set instead $\mstare(r)=M_{bulge}(r)$ in \Eq\ref{eq:massbudget}. The latter approximation stems from the fact that in most cases the bulge is significantly more compact than the disk implying that the gravitational effect of the disk on the bulge is negligible at the scales of interest to this work. The dynamical effect of the disk is then added in quadrature via the circular velocity, as described in \Sec\ref{subsec:disks}. We assume throughout that $M_{gas}(r)$ and $M_{bh}$ have a negligible contribution to the velocity dispersion within the central regions of the galaxy, and indeed we checked that including them using common prescriptions as suggested in the relevant literature \citep[e.g.,][]{MamonLokas2005,PeeplesShankar,Shankar2016}, does not alter any of our main results to any significant degree. We further discuss the impact of black hole mass to the central velocity dispersion in \Fig\ref{fig:Response} and of gas mass on the evolution of \sigmazM\ in \Sec\ref{Results-High}.

\subsection{Stellar Mass-to-light ratios}
\label{subsec:IMFr}

The transition from stellar light to stellar mass depends on the chosen mass-to-light ratios $\Upsilon_{0}=\mstare/L$ which is tightly linked to the chosen IMF. To specifically explore the impact of a scale-dependent IMF on our predicted velocity dispersions and their evolution with redshift, we follow \citet{Bernardi2018} and define a scale-dependent mass-to-light ratio $\Upsilon(R)$ of the type 
\begin{equation}
    \Upsilon(R) = 
    \begin{cases}
    \Upsilon_{0} \left( 1 + \phi - \xi\frac{R}{R_e}\right) & R < R_e \\
    \Upsilon_{0} & R \ge R_e
    \end{cases}
    \label{eq:ML}
\end{equation}
where, for simplicity, we set $\xi=\phi$, thus $\phi$ becomes the only parameter controlling the gradient of the profile. \Eq\ref{eq:ML} states that when $R\!\ge\!R_e$, the projected stellar mass-to-light ratio reduces to the usual constant stellar mass-to-light ratio $\Upsilon_{0}$, whereas at $R\!<\!Re$ the stellar mass-to-light ratio increases linearly until it reaches a maximum $\Upsilon_{max}=\Upsilon_0(1+\phi)$ at $R=0$. In Appendix 3, we include a Table containing values for $\Upsilon_{0}$ and $\Upsilon_{max}$ as a function of the measured galaxy velocity dispersion and $r$-band absolute magnitude, as calibrated by \citet{Bernardi19} and \citet{Dominguez19}, which we use to calculate stellar masses in the MaNGA sample when adopting a scale-dependent $\mstare/L$. Although their $\mstare/L$-values include the effects of gradients in age, metallicity, etc. as well, it is the IMF-gradient which matters most, so we simply refer to \Eq\ref{eq:ML} as describing an IMF-driven $\mstare/L$ gradient.

We can now directly multiply \Eq\ref{eq:ML} by the projected brightness profile (\Eq\ref{eq:sersic}) to obtain the new projected stellar mass density profile 
\begin{equation}
J(R) \equiv I(R)\Upsilon(R)\, .    
\end{equation}
The corresponding deprojected stellar mass density profile can then be retrieved from the integral \citep{Binney1982}
\begin{equation}
    \rho^*(r) = -\frac{1}{\pi}\int_r^\infty dR \frac{dJ/dR}{\sqrt{R^2 - r^2}} \, .
    \label{Deprojection}
\end{equation}
The 3D stellar mass density in \Eq\ref{Deprojection} can be written as a sum of two terms
\begin{equation}
    \rho^*(r) = \rho^*_{Ser}(r) + \Delta \rho^*(r)\, ,
\end{equation}
where the first term represents the standard deprojected stellar density profile assuming a constant mass-to-light ratio, $\rho^*_{Ser}(r) = \Upsilon_{0}\,\ell(r)$, and the second term represents the ``extra'' contribution to the stellar mass density profile. 
The latter can be computed as follows. The gradient in \Eq\ref{eq:ML} effectively adds, at each radius $R<R_e$, a projected stellar mass $\Delta \Upsilon(R)=\Upsilon_{0} \left(\phi - \phi\frac{R}{R_e}\right)$, which can be deprojected via \Eq\ref{Deprojection} to yield, when adopting a \Sersic light profile as in \Eq\ref{eq:sersic}, the extra stellar mass density 
\begin{equation}
\begin{split}
    \Delta \rho^*(r) = \frac{I_e \Upsilon_{0}}{\pi R_en}\int_y^1 \frac{dY}{Y\sqrt{Y^2 - y^2}} \exp\left(-b_n \left[ Y^{1/n}-1\right]\right) \times \\ 
    \left[n \phi Y + b_n(\phi -\phi Y)Y^{1/n}\right] \, ,
\end{split}
\end{equation}
where $Y=R/R_e$ and $y=r/R_e$. Once the deprojected stellar mass density $\rho^*(r)$ is acquired, the cumulative stellar mass $\mstare$ can be obtained from a simple spherical integral.

We conclude this \Sec pointing out that so far, when computing the effect of a scale-dependent $\mstare/L$ on velocity dispersion, we have strictly followed the formulation put forward by \citet{Bernardi2018}, \citet{Bernardi19} and \citet{Dominguez19}, i.e., we neglected any possible segregation in the phase-space distribution function of low- and high-mass stars \citep[e.g.,][]{Caravita21}. Nevertheless, as further detailed in Appendix 4, accounting for such a segregation in the Jeans Equation leads to relatively small differences in the shape and amplitude of \sigmal\ at $R<Re$. The main point is that, irrespective of the exact dynamical modelling applied via the Jeans Equation, whenever a realistic IMF-driven gradient is included in the $\mstare/L$, the resulting velocity dispersion profile in the inner regions is rather different from the one obtained with a constant $\mstare/L$, as emphasized by \citet[][]{Bernardi2018}.

\subsection{The contribution from the disk component}
\label{subsec:disks}

In the models in which we adopt a \Sersic+Exponential stellar light profile, the contribution from the galaxy disk to the aperture velocity dispersion $\sigma_{ap}$ is constructed as follows:
\begin{equation}
\begin{split}
    \sigma_{ap, disk}^2 \equiv \sigma_{disk}(R_{ap})^2 =\\ \frac{1}{M_{disk}(R_{ap})} \int_0^{R_{ap}} 2 \pi R \: \Sigma_{disk}(R) v_{rot}^2(R) sin^2 i\: dR \, .
\end{split}
\end{equation}
Here $M_{disk}=\Upsilon_0 L_{disk}$ is the stellar mass of the disk always computed assuming a constant mass-to-light ratio.
$R_{ap}$ is the aperture radius, $\Sigma(R)$ is the projected disk stellar mass density, $v_{rot}(R)$ the disk circular velocity, and $i$ is the disk inclination. We assume exponential disks of the form
\begin{equation}
    \Sigma_{disk}(R) = \frac{M_{disk}}{2 \pi h^2} \exp{(-R/h)}
\end{equation}
where $h$ is the disk scale length. The disk circular velocity is parameterised as follows (see e.g. \cite{Tonini2006})
\begin{equation}
v_{rot}(r)^2 = \frac{G M_{disk}}{2h}q^2 B\left(\frac{q}{2}\right) + v_{DM}^2 + \frac{GM_{bulge}(<r)}{r}
\end{equation}
where $q = r/h$, and $B(x) = I_0K_0 - I_1K_1$, a combination of modified Bessel Functions that account for the disk asphericity. The second term is the circular velocity of the dark matter halo, and the final term represents the gravitational contribution of the bulge on the disk.

\subsection{Effective radii and \Sersic\ indices}

Important features to properly describe the structural properties of bulges and spheroids are effective radii and \Sersic\ indices.
All our models start at $z=0$ from MaNGA galaxies with already measured structural (from light profile) parameters for both the bulge and disk components. To provide a complete prediction of velocity dispersion evolution along cosmic time, we would need reliable structural parameters even at higher redshifts. However, this information becomes gradually less clear or less accessible at earlier times on a galaxy-by-galaxy basis as in the MaNGA sample, but it can be retrieved when averaged in stellar/luminosity bins \citep[e.g.,][]{Dimauro19}. For this reason, we rely on empirically motivated analytic fits to the mean redshift evolution in effective radius $R_e$ and \Sersic index $n$ as a function of stellar mass as derived from a compilation of different data sets by \citet[][see also,e.g., \citealt{HuertasCompany2013}]{RicarteNat2018}. In the Equations that follow below, stellar mass is strictly defined via a constant mass-to-light ratio, i.e. $\mstare=\Upsilon_0 L$ with $L$ the total galaxy luminosity.

\begin{figure}
	\includegraphics[width=\columnwidth]{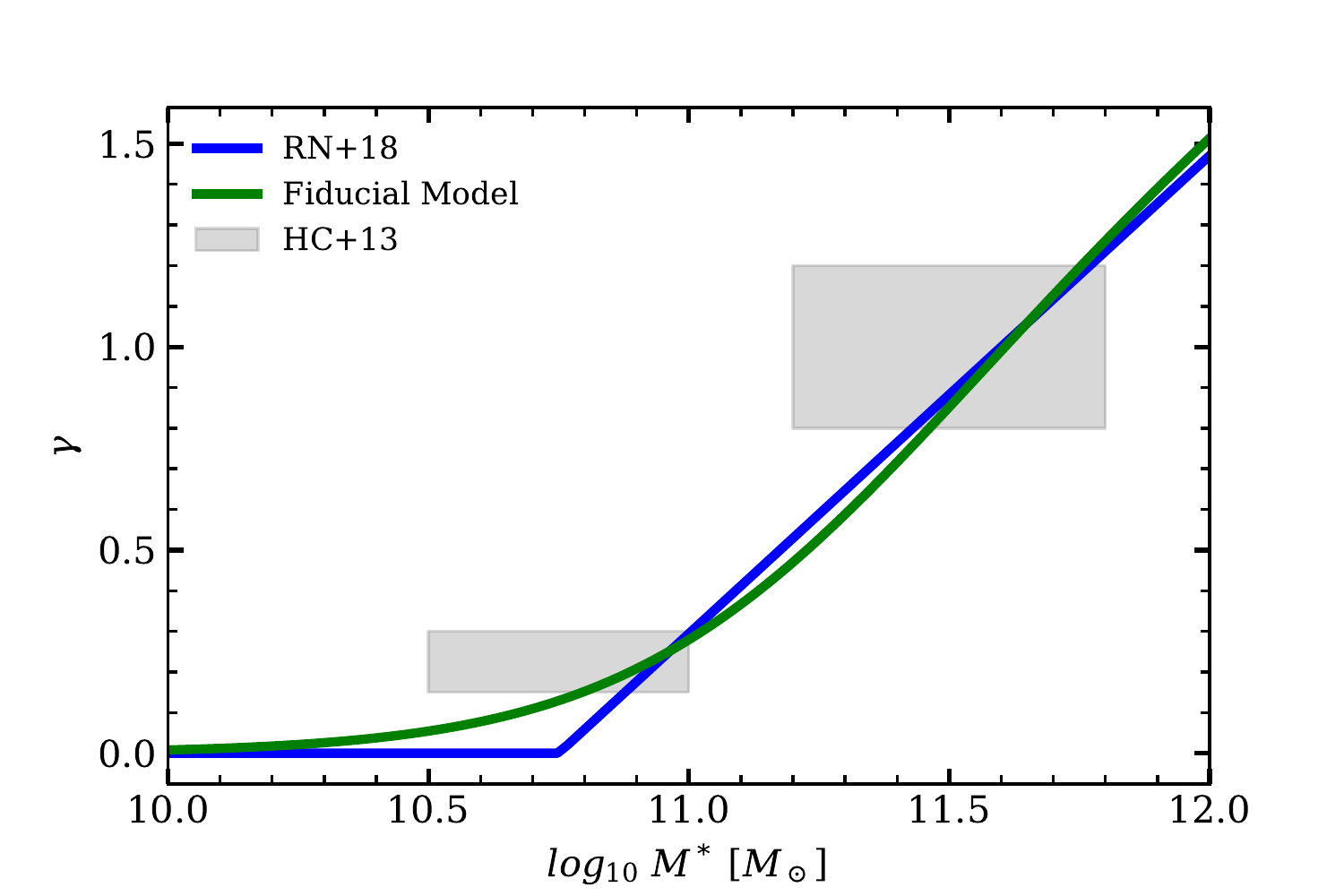}
    \caption{The effective radius evolution parameter $\gamma$ versus stellar mass. The shaded regions mark the regions of observationally calibrated $\gamma$ by \protect\cite{HuertasCompany2013} in some bins of stellar mass between $0\!<\!z\!<\!1$. The blue line shows the prescription used by \protect\cite{RicarteNat2018}. The green line shows the double power law fit utilized in this paper.}
    \label{fig:Gamma}
\end{figure}

Following \citet{RicarteNat2018}, we assume a stellar mass-dependent size evolution of the type  
\begin{equation}
    R_e(\mstare, z) = R_e(\mstare, 0)f(\mstare, z) \, .
    \label{eq:rez}
\end{equation}
The $R_e(\mstare, 0)$ are the effective radii as measured in the MaNGA sample, while the evolutionary factor $f(\mstare, z)$ is expressed as
\begin{equation}
    f(\mstare, z) = (1+z)^{-\gamma(\mstare)}\, .
    \label{eq:fmz}
\end{equation}
\citet{RicarteNat2018} assume $\gamma$ to be dependent on the galaxy stellar mass at $z=0$ and to evolve with redshift as
\begin{equation}
    \label{eq:RNgamma}
    \gamma(\mstare) = {\rm max}\left[0, \frac{1}{0.85}\left(log_{10}\, \mstare - 10.75\right)\right] \, .   
\end{equation}
\Eq\ref{eq:RNgamma} presents a discontinuity, which would propagate creating a visible break in velocity dispersion histories. We thus adopt a smoother dependence for $\gamma$ on redshift 
\begin{equation}
    \gamma(\mstare) = A\log_{10}\, \mstare 
    \left[ \left(B\log_{10}\, \mstare \right)^p + \left(C\log_{10}\, \mstare \right)^s \right]^{-1}
    \, ,
    \label{eq:gammaz}
\end{equation}
where $A = 3.05\times10^{-3}$, $B = 9.67\times10^{-2}$, $C = 0.204$, $p = -39.0$ and $s = -4.30$.
\Fig\ref{fig:Gamma} shows that our prescription is very similar to the one adopted by \citet{RicarteNat2018}, but allowing for a smooth variation in $\gamma$ and retaining consistency with some of the available observations \citep{HuertasCompany2013}. 

In a similar fashion to effective radii, \Sersic\ indices are assigned at any given epoch using a redshift evolution of the type
\begin{equation}
    n(\mstare, z) = n(\mstare, 0)(1+z)^{-1} \, ,
    \label{eq:sersicz}
\end{equation}
as suggested by \citet[][see also \citealt{Shankar17gamma}]{vanDokkum2010}. 
The local $n(\mstare, 0)$ are the \Sersic\ indices as measured in the MaNGA sample.

\subsection{Implementation}

The contributions from the bulge and disk (where present) are combined using \Eq\ref{combined}. The list of the parameters required by our model are shown in Table \ref{free}. To maintain excellent performance, we implemented our code in the C and C++ programming languages with Adaptive Richardson Extrapolation \citep{richardson1911} for fast and accurate integration, and paralleled using OpenMP. Our code is available \url{github.com/ChrisMarsden833/VelocityDispersion}.
We also provide in Appendix~\ref{AppA} a useful analytic formula to compute from our fiducial model with constant mass-to-light ratio the virial coefficient $\mathcal{F}=GM_{tot}(<\!R_e)/R_e\sigma_{ap}^2$, which in turn can be used to retrieve the dynamical mass, at any redshift $z \lesssim 3$, any aperture $R \lesssim R_e$, and effective radius and \Sersic\ index within the ranges probed in this work.

\begin{table}
 
	\centering
	\caption{List of variables used in our model.}
   \label{free}
   \label{tab:example_table}
    \begin{tabular}{rll}
    \hline
    Component & Variable & Description\\
    \hline
    &  $R_{ap}$  & Aperture \\
    \multirow{4}{*}{Bulge, Const. $\mstare/L$  $\begin{dcases*} \\ \\ \\ \end{dcases*}$}  & $M_{bulge}$  & Bulge Stellar Mass ($\Upsilon_0L_{bulge}$) \\
    &  $R_e$  & Half Mass Radius of the Bulge \\ &
    $\beta$ & Bulge Anisotropy Parameter \\ & $n$ & Sersic Index \\
    
    \multirow{6}{*}{Bulge, Var. $\mstare/L$ $\begin{dcases*} \\ \\ \\ \\ \\ \end{dcases*}$}  & $L_{bulge}$  & Bulge Luminosity \\
    &  $R_e$  & Half Mass Radius of the Bulge \\ &
    $\beta$ & Bulge Anisotropy Parameter \\ & $n$ & Sersic Index \\ & $\Upsilon_0$ & Mass-to-light ratio at $R_e$ \\ & $\phi$ & Mass-to-light ratio gradient \\
    
    \multirow{3}{*}{Disk $\begin{dcases*} \\ \\ \end{dcases*}$}  & $M_{disk}$  & Disk Stellar Mass ($\Upsilon_0L_{disk}$) \\
    &  $i$  & inclination \\ &
    $h$ & Disk Scale length \\
    
    \multirow{2}{*}{Halo $\begin{dcases*} \\ \end{dcases*}$}  & $M_{halo}$  & Halo Mass \\
    &  $c$  & Halo Concentration \\ 
    
    Black Hole \ \  & $M_{BH}$ & Black Hole Mass \\
    
     \hline
    \end{tabular}
\end{table}

\begin{figure*}
	\includegraphics[width=\textwidth]{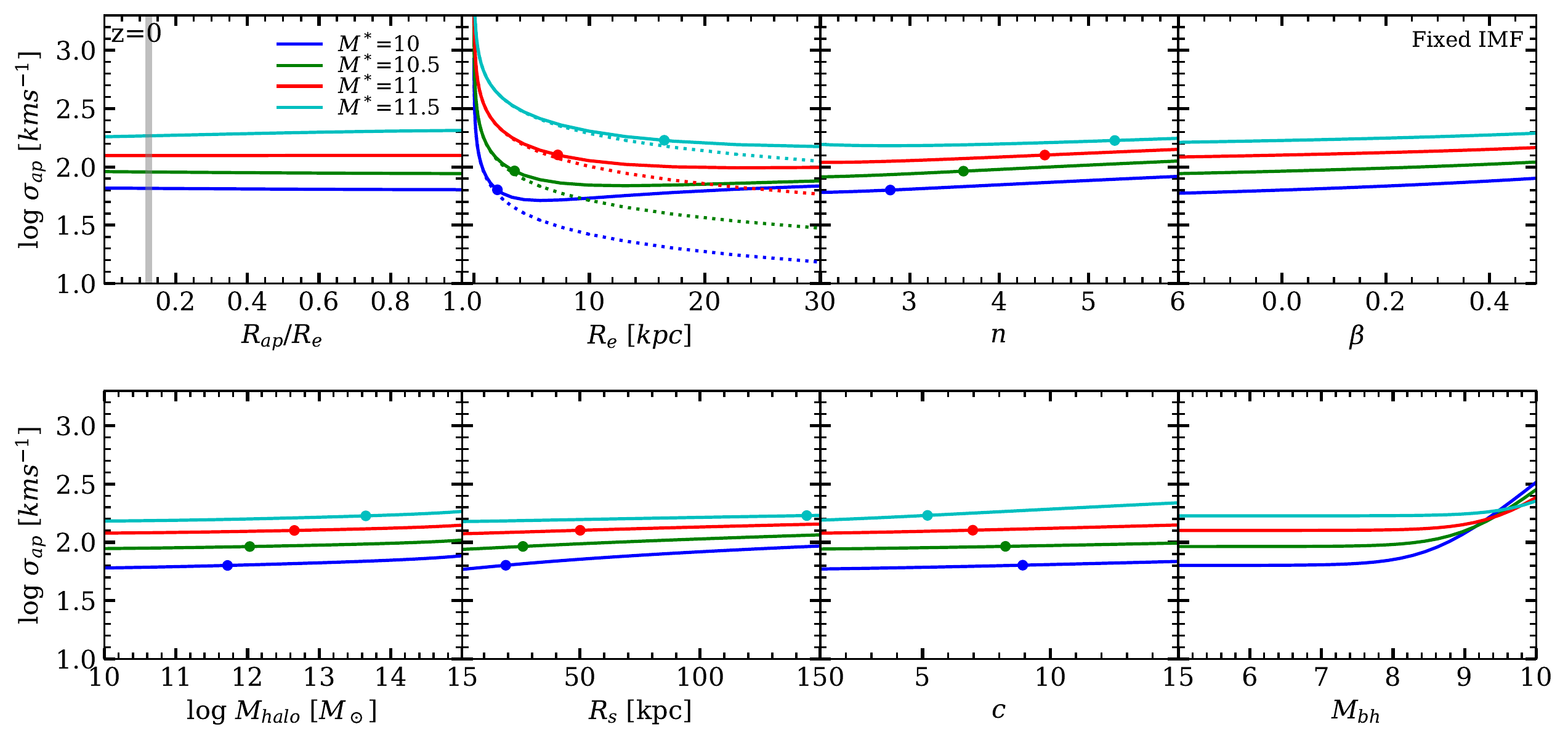}
    \caption{Dependence of (single-\Sersic\ profile) velocity dispersion $\sigma_{ap}$ on all main input parameters. Each coloured line represents a galaxy of fixed stellar mass at $z=0$, with all non-varying parameters in each panel assigned following the fiducial model (The aperture size is fixed at $R_e/8$, except in panel 1 where the response to aperture variation is shown. In this panel, the fiducial aperture is shown by the light grey vertical line). Solid circles show the typical values of some galaxy parameters for the chosen stellar mass bins (each respective parameter is fixed at these values in the panels when not the independent variable). Dotted lines show the calculated velocity dispersion when the dark matter component is neglected; except for the largest masses, the dark matter contribution within $R_e$ is negligible.}
    \label{fig:Response}
\end{figure*}

\section{Results} \label{Results}

In this \Sec we present the main outputs of our modelling of velocity dispersion detailed in \Sec\ref{Method}. We will first show in \Sec\ref{Results-Low} that our models provide a good match to the data from MaNGA in the local Universe. Anchoring our model at $z=0$ allows us to make predictions at higher redshifts, which we present in  \Sec\ref{Results-High}, based on empirically informed recipes for the evolution of the stellar mass and structural parameters of galaxies with redshift. For each of our most relevant steps, we will present results adopting both a constant and scale-dependent mass-to-light ratio.

\subsection{The Low Redshift Universe}
\label{Results-Low}

Before providing a detailed comparison of our models with the low redshift data, it is informative to first describe the overall behaviour of our Jeans modelling against all the main input parameters (\Fig\ref{fig:Response}). Here we simply show the case with a constant $\mstare/L$, as labelled, though the trends are similar for a radially-dependent mass-to-light ratio.  
The upper left panel of \Fig\ref{fig:Response} shows the velocity dispersion \sigmap\ as a function of aperture\footnote{The profile for $\sigmape$ as a function of aperture is consistent with the usual formula (e.g., \Eq A2 in \citealt{Shankar2019selectioneff}) for transforming $\sigmape$ between different aperture sizes, and also consistent with e.g. Figure 10 of \cite{FalconBarroso2017}} normalised to the effective radius $R_e$, whilst the other panels show the dependence of \sigmap, calculated within an aperture of $R=R_e/8$, on several different quantities, from left to right and from bottom to top, the effective radius, the \Sersic\ index $n$, the orbital anisotropy parameter $\beta$, the host halo mass $M_h$, the host halo scale radius $R_s$, the host halo concentration $c$, the central black hole mass $M_{bh}$. The different coloured lines are for different galaxy stellar masses, as labelled in the top, left panel. The solid circles in different panels mark the mean velocity dispersion in MaNGA that corresponds to that stellar mass. The most striking feature apparent from all panels is that our models predict velocity dispersions with extremely weak dependence on almost all input parameters, with only two possible exceptions, very small effective radii (upper middle left panel) and/or very large black hole masses (bottom right panel). The flatness of these curves strongly suggests that varying, e.g., the mapping between stellar mass and halo mass, and/or details on the structure of the stellar or the dark matter component would not significantly alter the velocity dispersion at fixed stellar mass (at least in the local Universe, we will see in the next section that this is not strictly true at higher redshifts). This weak dependence on the input parameters could in turn explain the tight correlation of velocity dispersion as a function of stellar mass, as anticipated by, e.g., \cite{Bernardi10}, and further discussed below.

Equipped with a clearer understanding of the dependencies of velocity dispersion on input parameters, we can now move to a close comparison to the observational data sets, most notably MaNGA (\Sec\ref{Data}). To this purpose, to each MaNGA galaxy with a measured velocity dispersion and luminosity profile, we assign a halo mass via abundance matching techniques. More specifically, we make use of
the \emph{inverse} of the stellar mass-halo mass relation (SMHM hereafter)\footnote{The latter is computed by first generating a large catalogue of host halo masses from the \cite{Tinker2010} halo mass function, assigning stellar masses via the (direct) stellar mass-halo mass relation, inclusive of normal scatter in stellar mass at fixed halo mass, then binning in stellar mass and finally computing the mean halo mass and scatter around the mean.}, and then apply the formalism discussed in \Sec\ref{Method} to predict its velocity dispersion within a given aperture \sigmap\ which we then compare with the measured one.
Here stellar mass is again intended to be the one computed at fixed mass-to-light ratio, $\mstare=\Upsilon_0L$. When adopting a gradient in the $\mstare/L$, we still assign halo masses to galaxies of a given luminosity $L$ using the \citet{Grylls2019STEEL} relation based on stellar masses derived from a constant $\Upsilon_0=\mstare/L$. In other words, a SMHM relation derived with constant $\mstare/L=\Upsilon_0$ can actually be considered, on average, a mapping between galaxy \emph{luminosity} and host halos mass, though the relation between stellar mass and halo mass when adopting a gradient will be different, in fact steeper, than the case with constant $\mstare/L$ \citep[see, e.g.,][]{Shankar17gamma}. 

\begin{figure}
	\includegraphics[width=\columnwidth]{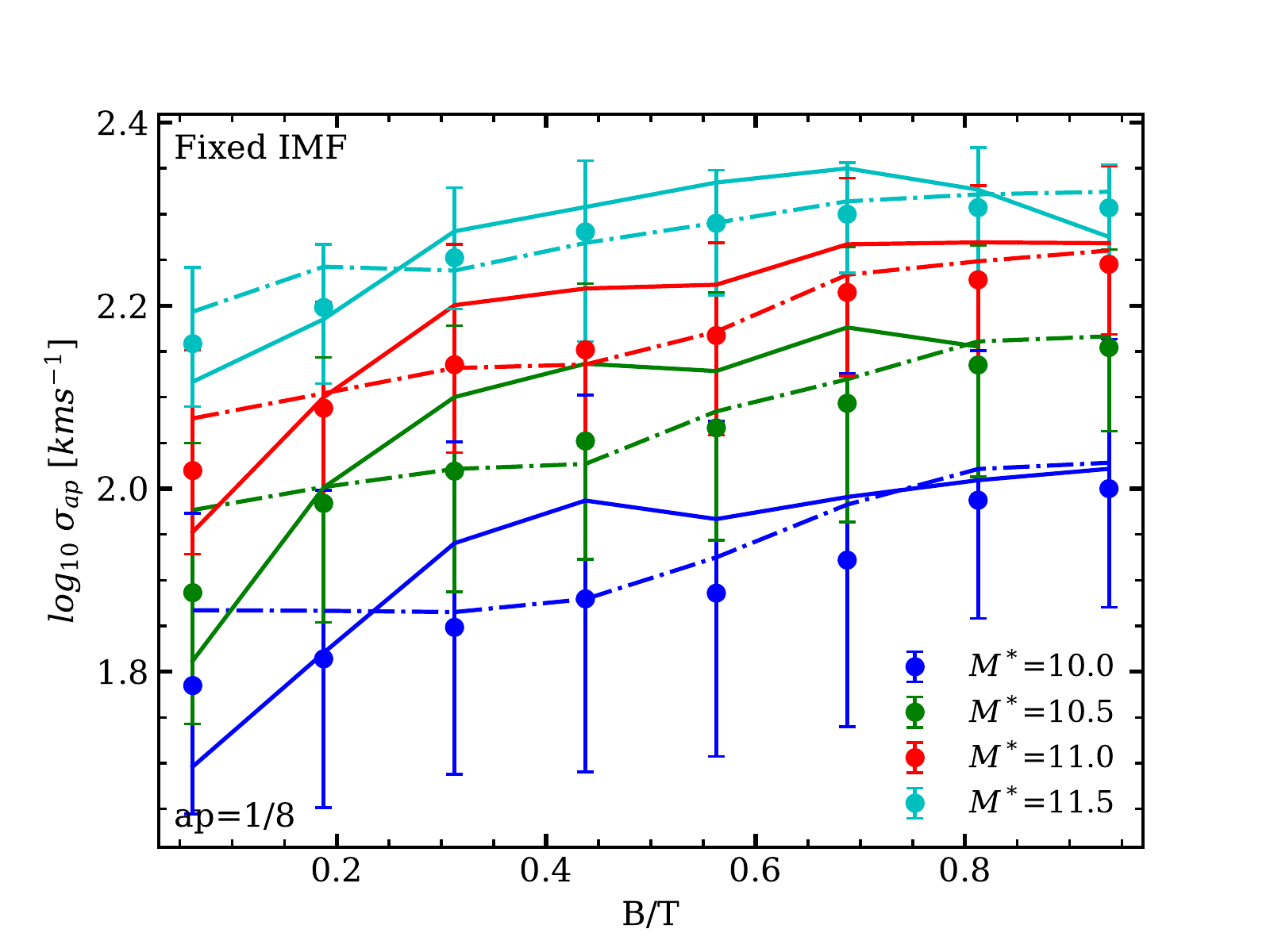}
     \caption{Coloured circles with error bars show the mean (with $1\sigma$ spread) $\sigma_{ap}$ versus $B/T$ for MANGa galaxies of different stellar mass of width 0.5 dex, as labelled. The solid and dot-dashed lines show the predictions of models with single-\Sersic\ and \Sersic+Exponential light profiles when we ignore $\mstare/L$ gradients.}
    \label{fig:BTsig}
\end{figure}

\Fig\ref{fig:BTsig} shows velocity dispersion computed within an aperture of $R_{ap}=R_e/8$ as a function of bulge-to-total (B/T) ratio as measured in the MaNGA sample (filled circles) for different stellar bins, as labelled, and assuming a constant $\mstare/L$. It is clear from the data that, for all stellar mass bins, velocity dispersion decreases as $B/T$ decreases, but it flattens out at $B/T \gtrsim 0.25$. This remains true for larger apertures $R\lesssim R_e$ and when we consider models with scale-dependent $\mstare/L$. We compare the MaNGA data with our predicted velocity dispersions -- computed following the methodology outlined in \Sec\ref{Method}, using both the full bulge-to-disk decomposition (solid lines) and the single \Sersic\ profile with zero anisotropy (dotted lines), assuming a constant $\mstare/L$.

The two component SerExp-based model predicts an initial increase of \sigmap\ with $B/T$ and then a flattening above $B/T \gtrsim 0.3$, while the single-\Sersic\ model is overall flatter, especially at low $B/T$, but still very much consistent with the data, implying that modelling velocity dispersion adopting spherical symmetry is still a good approximation for a wide range of galaxy morphologies, as long as appropriate galaxy luminosity-dependent effective radii and \Sersic\ indices are included in the model. In our models, the weak dependence of velocity dispersion on $B/T$ is just a reflection of the virial theorem -- velocity dispersion depends more on the total mass within an aperture and less on how this mass is distributed in it.  This may partly explain the similarity in the scaling relations of early and late-type galaxies \citep[e.g.,][]{Bernardi11,Ferrero21}, as well as the universality of the $M_{bh}-\sigma$ relation \citep[e.g.,][]{Ferrarese2000,Shankar2016}.

\begin{figure}
    \label{SigmaR}
	\includegraphics[width=\columnwidth]{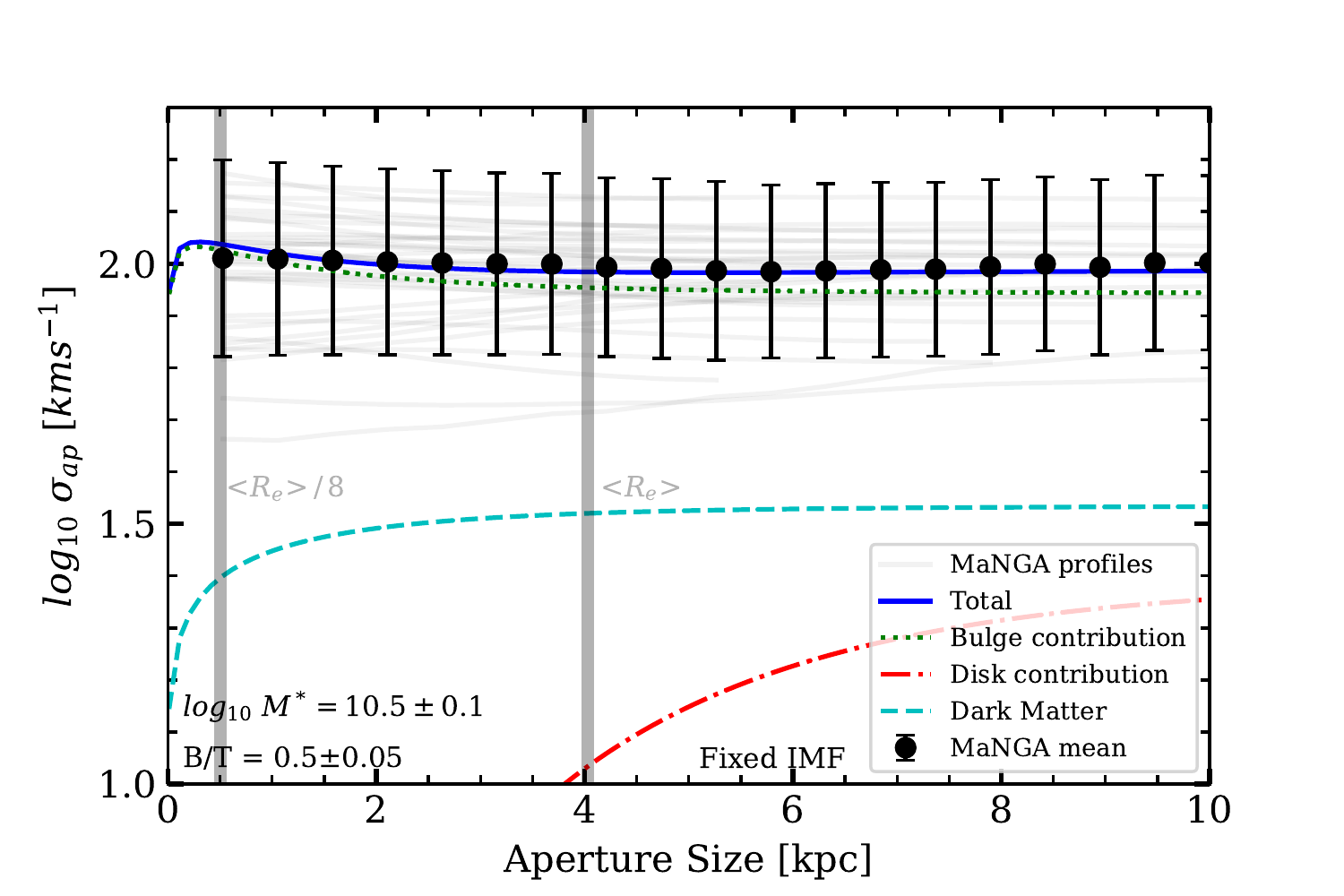}
     \caption{$\sigma_{ap}$ as a function of aperture size for a subset of galaxies within the MaNGA sample (faint grey lines, solid markers with error bars show the mean and $1\sigma$ spread), compared to the theoretical prediction (blue line) of our model for a galaxy with properties equal to the average properties of the galaxies from the selected MaNGA sample. Also shown are the corresponding bulge, disk and dark matter contributions to the total $\sigma_{ap}(R)$.}
    \label{fig:IndirectSim}
\end{figure}

\begin{figure*}
	\includegraphics[width=\textwidth]{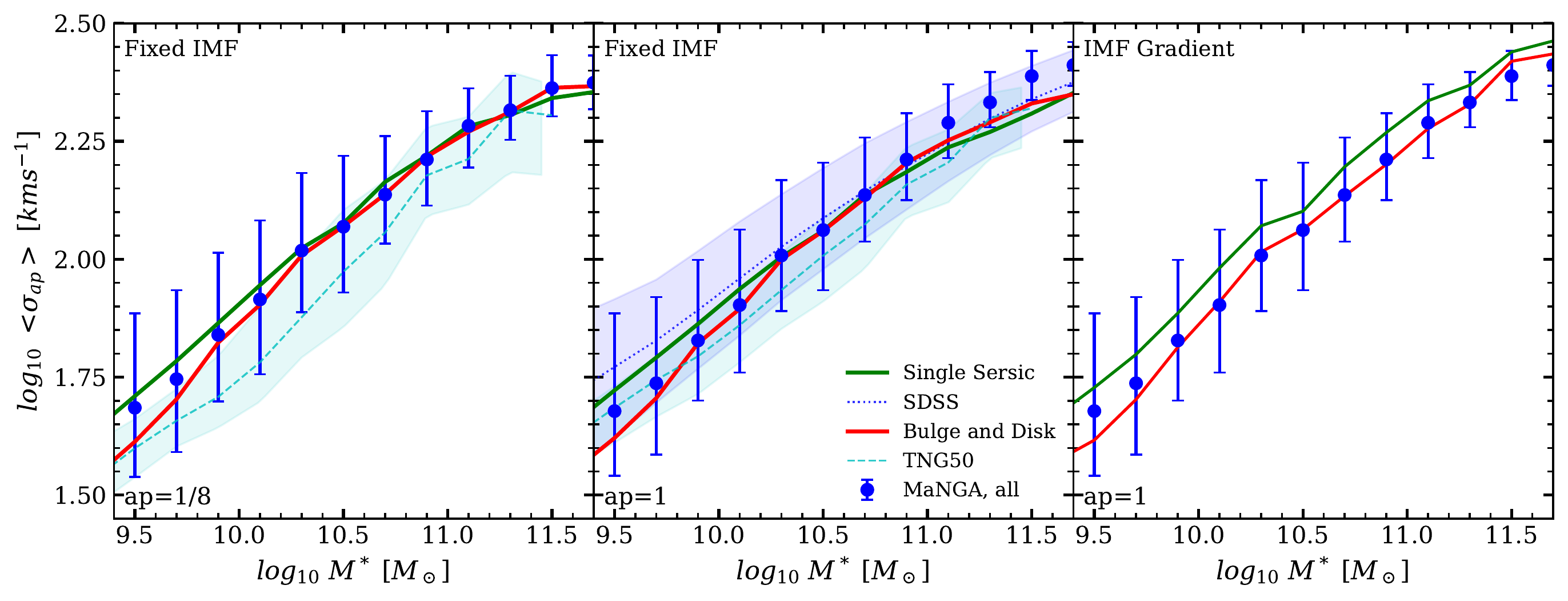}
    \caption{Predicted Faber-Jackson relation, $\sigma_{ap}$ versus stellar mass, for a single-\Sersic\ and two-component \Sersic+Exponential light profiles (green and red solid lines) with constant (left and middle panels) and scale-dependent (right panel) $\mstare/L$.  Blue filled circles show the mean FJ relation measured in MaNGA (in all three panels, the x-axis uses the \protect\cite{Mendel14} stellar masses), blue dotted line shows that in the SDSS and cyan dashed lines show the relation in the TNG50 simulation.}
    \label{fig:FJz0}
\end{figure*}

The main reason behind the weak dependence of velocity dispersion on $B/T$ in \Fig\ref{fig:BTsig} can be ascribed to the dominance of the bulge component to the dynamics in the inner regions. We would expect a progressively larger contribution of the disk when considering larger apertures. \Fig\ref{fig:IndirectSim} compares the (mean) predicted velocity dispersion \sigmap\ as a function of aperture size for galaxies with $B/T = 0.50 \pm 0.05$, and $\log_{10}\, \mstare/M_\odot = 10.5 \pm 0.1$ dex (solid blue line), with the mean velocity dispersion profile (solid circles) extracted from 25 galaxies (light gray lines) in the MaNGA sample in the same range of stellar and bulge masses (constant $\mstare/L$, $log_{10}\mstare=10.5\pm0.1$ and $B/T\pm0.05$). The model very well aligns with the data (solid blue line), and it shows that the total velocity dispersion is vastly dominated by the bulge (dotted green line). The relative dynamical contributions of the disk itself on the rotational velocity and thus on the velocity dispersion (dot-dashed red line) is relatively minor even up to the effective radius ($\lesssim 40\%$ at $R=R_e$). Similarly, even the contribution of the dark matter component (dashed cyan line) is contained in the inner regions, reaching $\sim 30\%$ at $R_e \sim 8$, though we will see dark matter fractions are relevant in shaping the evolution of velocity dispersion with cosmic time. Once again, both data and models in \Fig\ref{fig:IndirectSim} highlight an extremely flat velocity dispersion as a function of aperture size.

To further validate our modelling, in \Fig \ref{fig:FJz0} we compare our predicted mean velocity dispersion-stellar mass relation, also known as the Faber-Jackson (FJ) relation \citep[][]{FaberJackson}, with the one measured in MaNGA (filled blue circles). The left panel shows that our predicted $\sigma_{ap}$ calculated at an aperture of $R=R_e/8$ for both models with a single-\Sersic and \Sersic+Exponential light profiles (solid green and red lines, respectively) and constant $\mstare/L$ are very similar to each other and to the MaNGA data.
This very good match with the data extends up to the effective radius, as shown in the middle panel of \Fig\ref{fig:FJz0}, in which we also include the FJ from the SDSS data\footnote{We apply a constant horizontal shift of -0.05 dex to the $\sigma_{ap}-\mstare$ SDSS relation by \citet{Shankar2019selectioneff} who increased all their stellar masses by an average 0.05 dex to account for the relatively small difference between the \citet{Mendel14} and \citet{Bell03} mass-to-light ratios following \citet{Bernardi17}.} as calibrated by \citet[][dotted blue line and purple region]{Shankar2019selectioneff}. Both data sets are in very good agreement with each other despite the significant differences in galaxy selections and velocity dispersion measurements. We note that all the velocity dispersions adopted to calibrate the FJ relations in \Fig\ref{fig:FJz0} include, as anticipated in \Sec\ref{Data}, the rotational component as measured in the MaNGA sample. We verified that neglecting the rotational component would yield very similar results on the FJ relation, in line with the results of our models presented in \Fig\ref{fig:IndirectSim}.

The right panel of \Fig\ref{fig:FJz0} shows the FJ relation at an aperture of $R=R_e$ with the same MaNGA data as before, only now the model curves result from including the IMF-driven $\mstare/L$ gradient as described in \Sec\ref{subsec:IMFr}. 

It is interesting to compare the curves in this panel with those in the middle one (the symbols are the same but the stellar masses somewhat different due to the inclusion of the IMF-driven $\mstare/L$ gradient).  We have already made the point that the two-component \Sersic+Exponential models (red) are more accurate.  In the middle panel, these models systematically underestimate the measurements at large masses; this is the difference that is usually explained by assuming that the IMF becomes bottom heavy at large masses.  In contrast, the panel on the right, which includes an IMF-driven $\mstare/L$ gradient in the Jeans equation analysis -- shows no such systematic underestimate.  Thus, the price we pay for ignoring IMF-related effects altogether is the small deficit between the red curves and the measurements in the middle panel.  

For completeness, in both the left and middle panels of \Fig \ref{fig:FJz0} we also include the FJ relation derived from the TNG50 simulation (we discuss below and in Appendix~\ref{AppB} how we extract velocity dispersions in TNG50), which is also remarkably close to the data at high masses and only slightly below them at $\log_{10}\, \mstare/M_\odot \lesssim 11$.

\begin{figure*}
	\includegraphics[width=\textwidth]{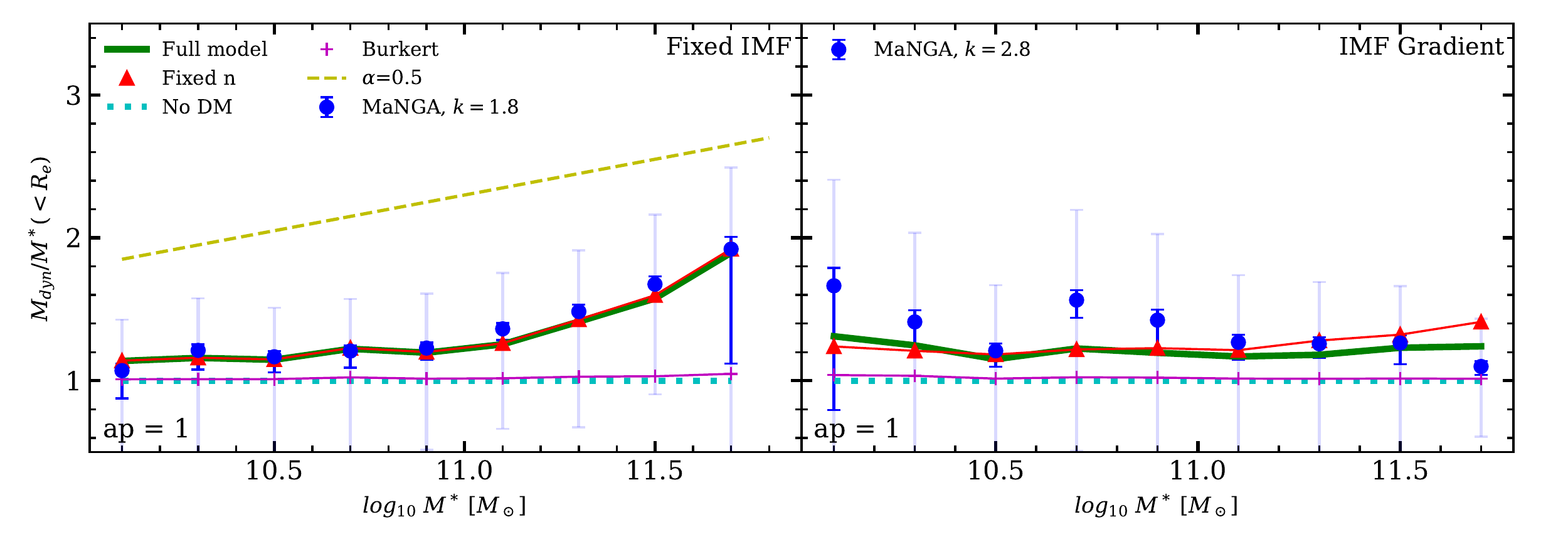}
    \caption{Dynamical-to-stellar mass ratio as a function of stellar mass for a model with constant (left) and scale-dependent (right) $\mstare/L$. Blue points represent mean values derived from the velocity dispersion from the MaNGA sample, with faint error bars showing the associated $1\sigma$ uncertainty, and solid bars showing the error on the mean. The green line, red triangles, cyan dotted line and purple crosses represent the predicted values from our fiducial model, our fiducial model at fixed \Sersic\ index, no dark matter and with a Burkert DM profile respectively. Accounting for an IMF-driven $\mstare/L$ gradient makes $M_{dyn}\propto \mstare$ within the effective radius. We impose a lower limit on these plots of $\mstare \ge 10^{10}\, M_\odot$, as below this threshold the data on $M_{dyn}\propto \mstare$ become much more noisy and less accurate.} The dashed yellow line in the left panel indicated a slope of $\alpha=0.5$ to guide the eye.
    \label{fig:FP}
\end{figure*}

\Fig\ref{fig:FP} shows with filled blue circles the ratio between dynamical mass and stellar mass both as measured in MaNGA within an aperture equal to the effective radius $R=R_e$ for a constant (left panel) and scale-dependent $\mstare/L$ (right panel) with superposed errors on the mean (thick lines) and variance (shaded lines). In this plot we retain all galaxies irrespective of their morphological types though, interestingly, very similar results are found when limiting the analysis to only early-type or elliptical galaxies.\footnote{In this plot we limit the analysis to galaxies with \Sersic\ index $n>1.8$ to avoid the inclusion of noisy velocity dispersions.} The dynamical masses are simply computed as
\begin{equation}
     M_{dyn} = k\sigma^2 R_e/G,
        \label{eq:dynamic}
\end{equation}

with $k=1.8$ and $2.8$ in the left and right panels (We note that \Eq\ref{eq:dynamic} is only required for the observed galaxies, as the dynamical masses for the galaxies that are directly modelled are known.). These constant ``virial'' coefficients $k$ are simply chosen to normalise the $M_{dyn}/\mstare$ ratio towards unity at low stellar masses, and thus differ from the more accurate values computed in Appendix~\ref{AppA}, which include the effects of, e.g., stellar profile and anisotropy. The increase in the normalization factor $k$ in \Eq\ref{eq:dynamic} when including the IMF-driven gradients simply reflects the fact that the stellar masses increase in the presence of gradients, and thus a proportionally higher $k$ is required to retrieve a similar $M_{dyn}/\mstare$ ratio of order unity. When assuming a constant IMF (left panel), the measurements (symbols) show evidence for a ``slope'' in the dynamical-to-stellar mass ratio (see \Sec\ref{sec:Intro}) especially at $R=R_e$.  (The slope is shallower at $R_e/8$.) This trend of increasing $M_{dyn}/\mstare$ with stellar mass is reproduced by our fixed IMF models, without any extra fine-tuning. In these models it is mainly the fraction of dark matter within the effective radius which drives this trend:  if we assume that there is no dark matter (dotted cyan line), then our models return no trend. In fact, switching to a \citet{Burkert95} dark matter profile generates a flat $M_{dyn}/\mstare$ ratio (purple line with cross markers). In addition, structural parameters tend to play a secondary role in shaping the $M_{dyn}/\mstare$ ratio.  For example, setting \Sersic\ index $n=4$ for all galaxies yields nearly identical results to our base model (red line with triangular markers). 

It is interesting that our fixed IMF model seems to be in good agreement with the data, despite the large body of literature arguing that the IMF must change across the population (if, as we have done here, one assumes there are no $\mstare/L$ gradients).  In fact, if we plot $M_{dyn}/\mstare$ versus $\sigma$ instead, then the models significantly underpredict the measurements as $\sigma$ increases, in good agreement with the literature \citep[e.g.][]{Cappellari13}.  The agreement in the left hand panels of Figure~\ref{fig:FP} must result from the fact that the scatter between $\sigma$ and $\mstare$ is large enough to hide the problems at large $\sigma$.

For completeness, the right panel of \Fig\ref{fig:FP} shows $M_{dyn}/\mstare$ after assuming an IMF-driven $\mstare/L$ gradient (this changes the denominator of the data, and both numerator and denominator of the model) as described in \Sec\ref{subsec:IMFr}. In this case the resulting $M_{dyn}/\mstare$ ratio is flat as a function of stellar mass both in the MaNGA data (blue filled circles) as well as in the models. (The ratio is also flat when plotted as a function of $\sigma$.)  Reducing the fraction of dark matter (purple line with cross markers) or setting $n=4$ for all objects (solid red line with triangles) has negligible impact on the predicted ratio. All in all, our results show that when an IMF-driven gradient in $\mstare/L$ is included in the Jeans analysis, then $M_{dyn}\propto \mstare$ within $R_e$, in agreement with \cite{Bernardi2020FP}. 

\subsection{The high redshift Universe} 
\label{Results-High}

\begin{figure*}
	\includegraphics[width=\textwidth]{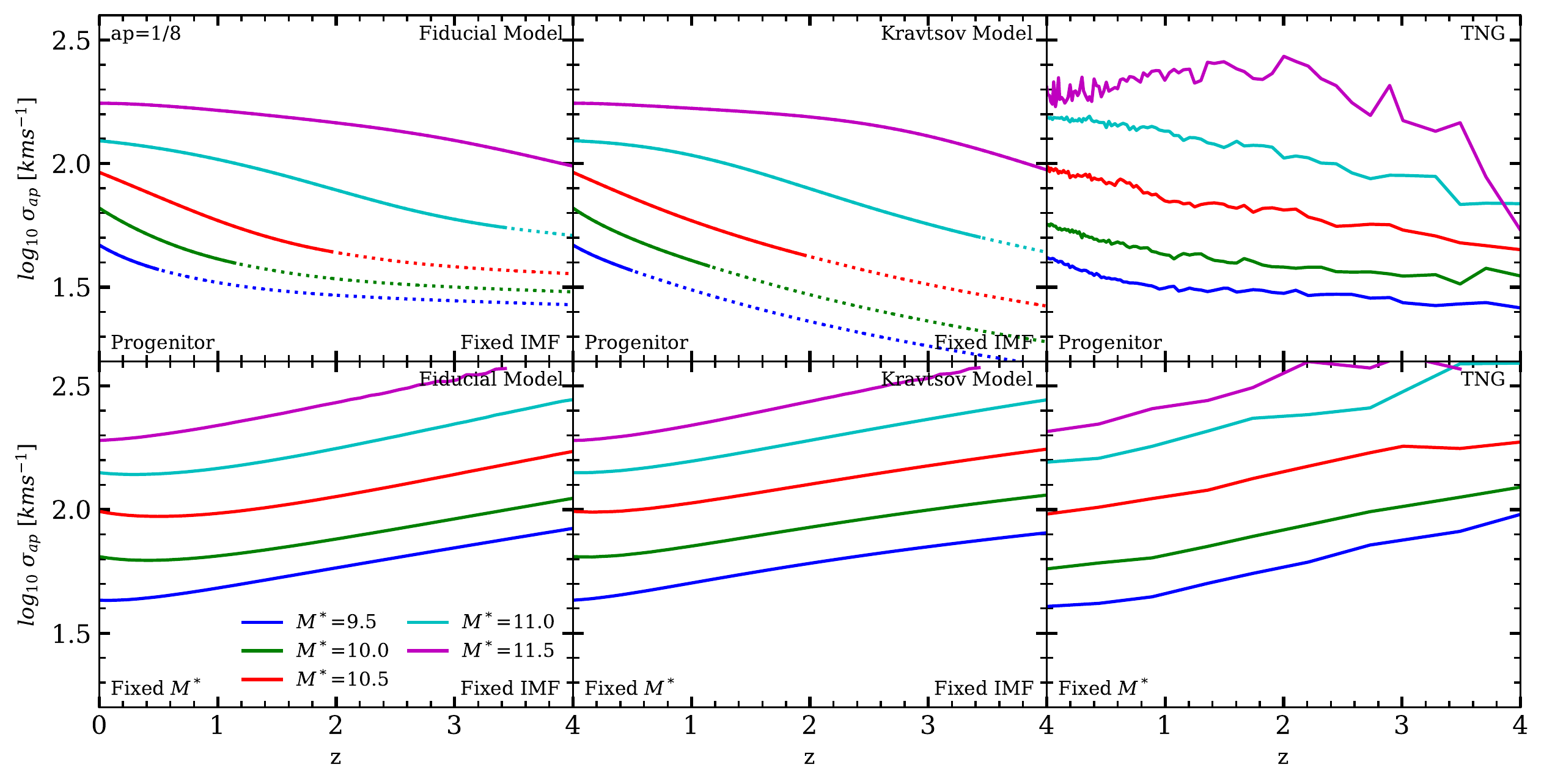}
    \caption{Predicted \sigmazM\ evolutionary tracks for models with constant $\mstare/L$ along the progenitors (top panels) and at fixed stellar mass (bottom panels) for different stellar masses, as labelled. The left column shows our fiducial model described in \Sec\ref{Method}. The central column shows the fiducial model with effective radii computed via the \protect\cite{Kravtsov2013} relation. The right column shows $\sigma_{ap}$ for galaxies within the same mass bins extracted from the TNG50 simulation.
    Dotted lines mark the region where the input empirical scaling relations fall outside the observational parameter space in which they were calibrated (see text for details). 
    }
    \label{fig:Size3x2}
\end{figure*}

We showed in \Sec\ref{Results-Low} that our model to build velocity dispersion profiles for galaxies of different stellar masses is successful in matching the local data. In particular, we showed that the models can reproduce the weak dependence on $B/T$ ratio, as well as the full velocity dispersion profile, FJ relation and dynamical-to-stellar mass ratio, irrespective of the exact (constant or spatially varying) $\mstare/L$. In this \Sec\ref{Results-High} we make predictions on the velocity dispersion \sigmazM\ evolutionary tracks, and their implications for the evolution of the FJ relation and of the dynamical-to-stellar mass ratio. To generate the \sigmazM\ tracks we adopt the following strategy:
\begin{itemize}
\item We consider stellar masses defined by a constant $\mstare/L$ in the MaNGA data set. We then select a stellar mass bin and compute its mean effective radius and \Sersic\ index which we choose as our starting point in \Eqs\ref{eq:rez} and \ref{eq:sersicz}. 
\item We compute the mean host halo mass competing to the chosen bin of galaxy stellar mass at $z=0$ via the inverse of the SMHM relation, and then follow backwards in time the mean halo assembly history $\langle M_{halo}(z)\rangle$ competing to that halo. We use the halo growth histories from \citet{vdb2014}, which were derived from detailed analytic recipes tested against N-body dark matter simulations. 
\item At each time step we then apply our numerical formalism (\Sec\ref{Method}) to derive the velocity dispersion. In our reference/fiducial model we adopt the \citet[][]{Grylls2019STEEL} (inverted) SMHM relation and assume a NFW profile for the dark matter component. 
\item We then evolve at each time step the (mean) effective radius via \Eqs\ref{eq:rez}, \ref{eq:fmz} and \ref{eq:gammaz}, and the \Sersic\ index via \Eq\ref{eq:sersicz}. 
\end{itemize}
We will further discuss below that the main trends in our output \sigmazM\ are qualitatively unchanged when varying, within reason, any of the input parameters or their exact evolution with redshift. In what follows, we rely only on single-\Sersic\ profiles for two reasons: 1) We do not need to assume any time evolution in the $B/T$ along the main progenitors, which is still debated (and in any case not too relevant to velocity dispersion as emphasized in, e.g., \Fig\ref{fig:BTsig}); 2) we decrease the number of parameters to only the redshift evolution in effective radius and \Sersic\ index. All the predictions on the \sigmazM\ presented in this \Sec\ref{Results-High} are calculated, unless otherwise noted, within an aperture of $R=R_e/8$.  Although $Re/8$ is difficult to resolve at higher $z$, this choice is mostly driven by the comparison with the TNG50 simulation. Increasing the aperture to $R=R_e$, for example, would significantly increase the computational cost of extracting and computing TNG50 velocity dispersions on a galaxy-by-galaxy basis. In addition, larger apertures tend to give more noisy results from TNG50. We note, however, that our predicted \sigmazM\ evolutionary tracks are very similar in shape when adopting an aperture equal to the effective radius, they only slightly reduce in normalization. As discussed in \Sec\ref{Results-Low}, the choice of a constant or scale-dependent $\mstare/L$ produces different outputs in the dynamical-to-stellar mass ratio within the effective radius (\Fig\ref{fig:FP}), which could be controlled by either the fraction of inner dark matter mass and/or by a larger amount of low-mass stars (bottom-heavy IMF, \Eq\ref{eq:ML}). Therefore, in what follows, we will present results on the evolution of \sigmazM\ with both a fixed IMF and IMF-gradient driven $\mstare/L[r]$. Since there is no consensus on how IMF-driven gradients evolve, we include them using a very simplified model:  we simply set $\Upsilon_0=4$ and $\phi=\xi=1$ in equation~(\ref{eq:ML}) for all the galaxies in our higher $z$ runs. 

\Fig\ref{fig:Size3x2} shows our predicted velocity dispersion evolutionary tracks \sigmazM\ at an aperture of $R_{ap}=R_e$, for different stellar masses, as labelled, along the main dark matter progenitors (top panels) and at fixed stellar mass (bottom panels). The left panels report the results of our fiducial model, the middle panels contain outputs of a model that replaces our empirical relations for effective radii (\Eq\ref{eq:rez}) with an empirical linear and tight relation between effective radius and halo virial radius $R_e\propto R_{vir}$ \citep[e.g.,][]{Kravtsov2013,Stringer14,Huang17,Zanisi20,Zanisi21}, and the right panels show the results extracted from the TNG50 hydrodynamical simulation, for the same bins of stellar mass, with stellar velocity dispersions computed as described in Appendix~2 and averaged within the same aperture as in our semi-empirical models. The point to note is that all models show a similar imprint of ``downsizing'' in the predicted \sigmazM, with the more massive galaxies approaching the local value of \sigmazM\ at earlier epochs, comparably to what observed in the relative mass growth histories of central galaxies along their main progenitor tracks \citep[e.g.,][]{Moster13,Behroozi19,Grylls2019STEEL,Shankar20}. \citet{RicarteNat2018} also found a clear sign of downsizing in their velocity dispersion evolutionary tracks. 

Interestingly, we do not find the downsizing in \sigmazM\ to be strongly dependent on the exact choice of our model for the effective radius. For example, computing the \sigmazM\ in our fiducial model assuming no redshift evolution in effective radii ($\gamma=0$ in \Eq\ref{eq:gammaz}), would yield similar results. In addition, the close similarity between the fiducial model and the Kravtsov model in the predicted \sigmazM\ tracks opens up the possibility of computing reliable galaxy velocity dispersion in semi-analytic and semi-empirical models via the use of only the virial radius of the host dark matter halo \citep[see also][]{Zanisi21}. The \sigmazM\ evolutionary tracks in \Fig\ref{fig:Size3x2} are shown down to a minimum stellar mass of $\log \mstare/M_\odot=9$, below this threshold we do not have data to calibrate the SMHM relation, nor sufficient velocity dispersion measurements in MaNGA, and also the $R_e \propto R_{vir}$ relation has not been well constrained below $R_e \lesssim 0.5$ kpc  \citep[e.g.,][]{Kravtsov2013,Somerville18}, which are the typical scales of galaxies with $\mstare \lesssim 10^9\, M_\odot$. We therefore mark with dotted lines in \Fig\ref{fig:Size3x2} all extrapolations in our data-driven models below this mass threshold of $\log_{10} \mstare/M_\odot=9$. 

\begin{figure}
	\includegraphics[width=\columnwidth]{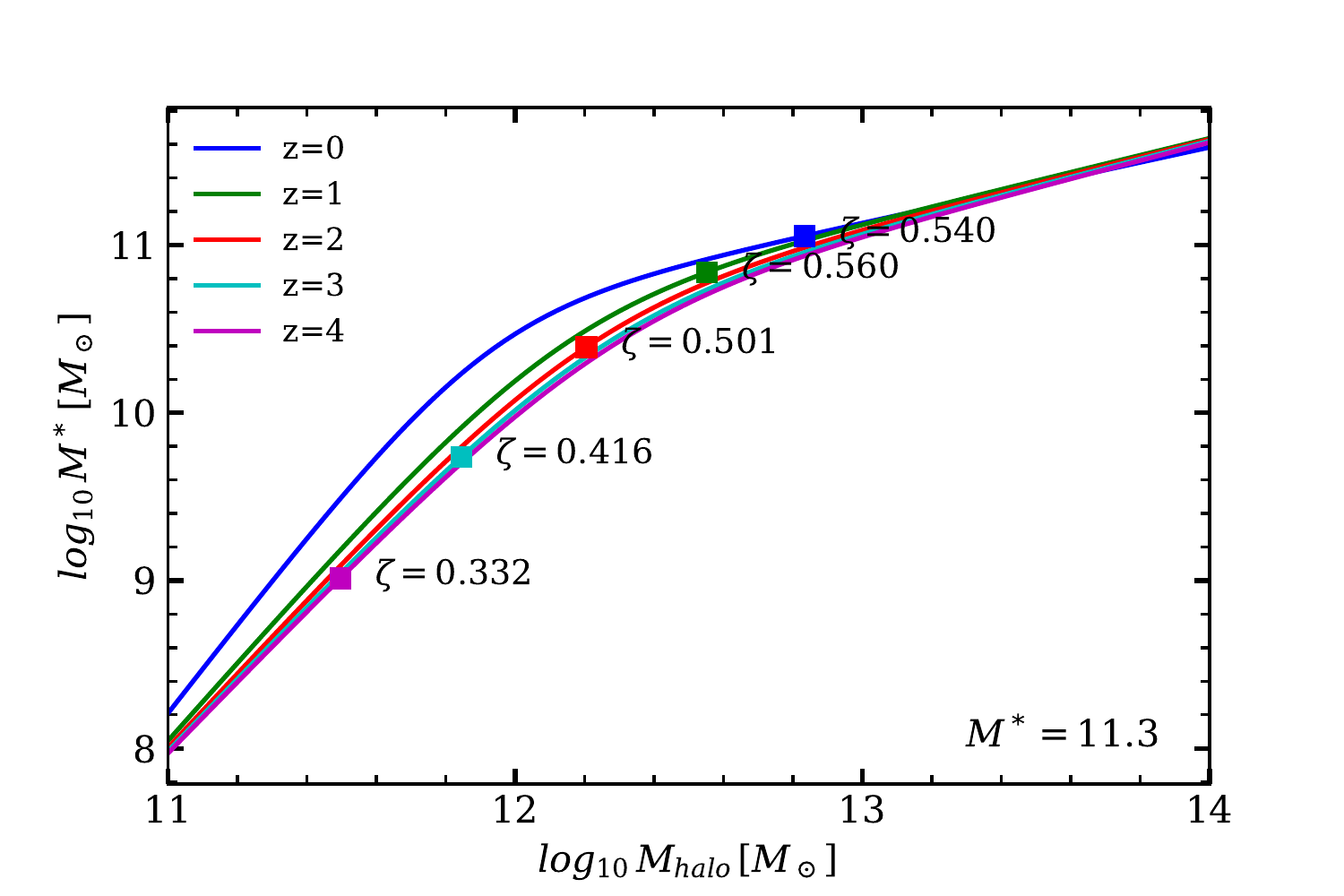}
    \caption{The \protect\cite{Grylls2019STEEL} stellar mass-halo mass (SMHM) relation at different redshifts, as labelled. The coloured squares mark the value of $\zeta$ for a galaxy of progenitor mass $log_{10}$ $\mstare=11.3$ $[M_\odot]$ at $z=0$, at the different redshift steps listed in the legend. The parameter $\zeta$ is the stellar-mass-to-total-mass ratio within $R_e$ ($\zeta = \mstare(<R_e)/(M_{halo}(<R_e) + \mstare(<R_e))$). $\zeta$ has a weak evolution above the knee but drops significantly when below the knee of the SMHM relation.}
    \label{fig:SMHM}
\end{figure}

\begin{figure*}
	\includegraphics[width=\textwidth]{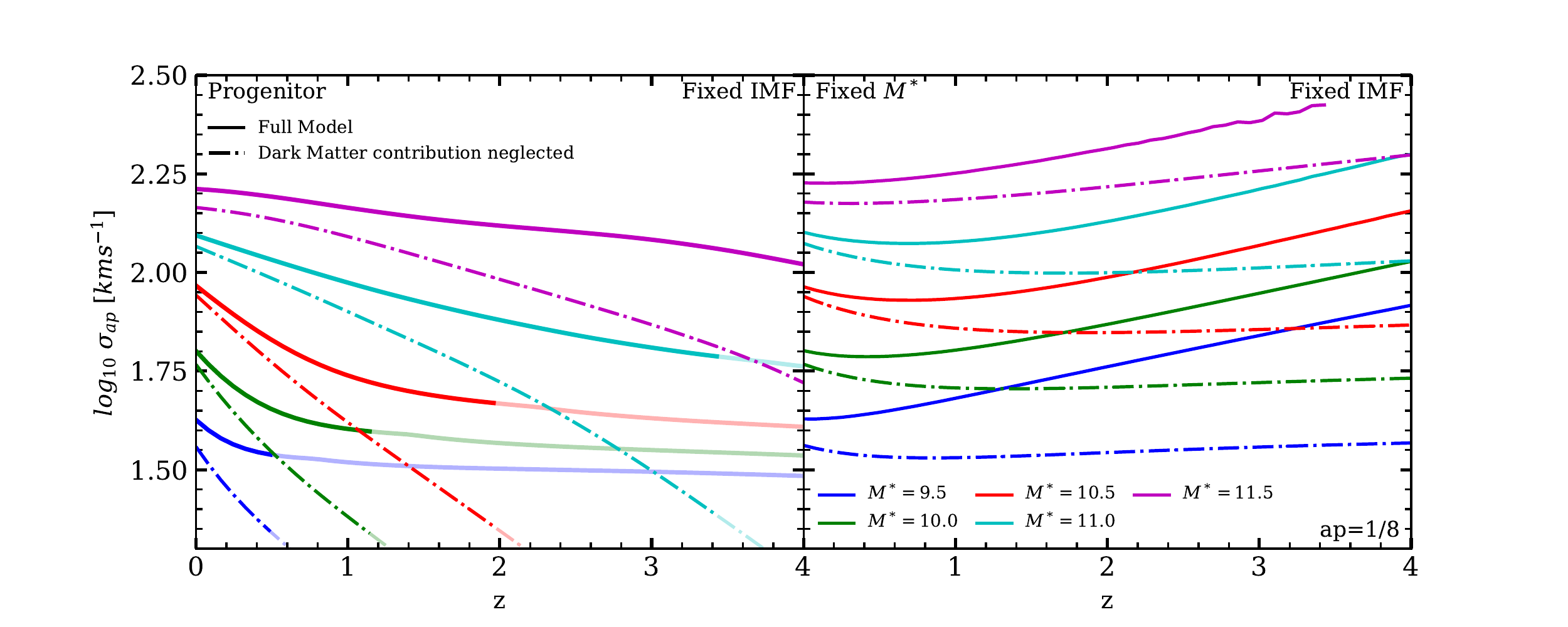}
    \caption{The average velocity dispersion for galaxies of fixed stellar mass (at $z=0$) vs redshift for a fixed IMF (and no $\mstare/L$ gradient). The solid lines show the fiducial model (identical to the left panels of figure \ref{fig:Size3x2}), whereas the dot-dashed lines are without the dark matter component. The faint lines show the extrapolations of these models in the regions where the input scaling relations exceed the parameter space of the MaNGA catalogue. Neglecting the dark matter component produces a much steeper evolution of $\sigma$ with cosmic time.}
    \label{fig:SigmaTracks}
\end{figure*}

Instead, the downsizing in \sigmazM\ appears to be largely driven by the shape of the input stellar mass-halo mass (SMHM) relation. The different lines in \Fig\ref{fig:SMHM} show the SMHM relation at different redshifts, as labelled, as derived by \citet{Grylls2019STEEL} and updated in \citet{Zanisi20}. The results show a weak evolution of the SMHM relation, especially at $z\gtrsim 1$ and at higher stellar masses, a conclusion shared by several previous works \citep[e.g.,][]{Shankar06,Moster2010}. The persistent double power-law shape of the SMHM relation inevitably implies a varying fraction of stellar mass across different halo masses \citep[e.g.,][]{Kravtsov04,Vale04,Shankar06}. In particular, a dark matter halo residing below the ``knee'' of the SMHM relation, which is around $M_h\sim 10^{12}\, \Msun$, tends to host, on average, a central galaxy with stellar mass relatively lower than the stellar mass of a central galaxy hosted in a more massive halo above the knee of the SMHM relation. In other words, the SMHM relation predicts that there are proportionally less stars formed in lower mass haloes \citep[e.g.,][]{Shankar06,MattJoop}.   

To illustrate this effect, we mark with filled squares the evolution along the SMHM relation of the ratio $\zeta=\mstare(R_e)/[\mstare(R_e)+M_{halo}(R_e)]$ for a galaxy with stellar mass of $\mstare=10^{11}\, M_\odot$ at $z=0$. It is apparent that, whilst the ratio $\zeta$ remains similar when the galaxy resides above the knee of the SMHM relation (in this specific case $\zeta \sim 0.55$), it rapidly drops (in this case by up to $\sim 40\%$ at $1<z<4$) when the galaxy crosses the knee of the SMHM relation. A similar behaviour in $\zeta$ is observed for all galaxy masses of interest here. When the stellar mass $\mstare(<R)$ drops sufficiently, a condition that is more easily met below the knee of the SMHM relation, the dark matter ``takes over'' in controlling the velocity dispersion within the aperture $R$. It is then natural to expect a flattening in the \sigmazM\ tracks as the evolution in $M_{halo}[<R_e,z]$ is relatively weak along the progenitor tracks. In summary, the top panels of \Fig\ref{fig:Size3x2} point to a pivotal role of the SMHM relation, and in particular of the location of its knee, in shaping the downsizing in the \sigmazM\ evolutionary tracks, at least in models with no $\mstare/L$ gradient. 

The bottom panels of \Fig\ref{fig:Size3x2} present the \sigmazM\ at \emph{fixed} stellar mass for the same models as in the corresponding top panels and for the same stellar masses at $z=0$. To generate these plots we build a mock catalogue of central galaxies and parent dark matter haloes at each redshift of interest and then bin in stellar mass. In this case the velocity dispersions are predicted to steadily increase at earlier epochs in a nearly parallel fashion. This behaviour could also be ascribed to the role of dark matter in the inner regions. Given the weak redshift evolution of the input SMHM relation \citep[e.g.,][]{Moster2010}, when moving backwards in time at fixed stellar mass implies retaining a similar host halo mass and thus an increasing central dark matter mass density induced by progressively smaller virial radii and increasing background densities. 

Our fixed IMF models (which ignore $\mstare/L$ gradients) predict an evolution of velocity dispersion at fixed stellar mass of the type \sigmazM$\propto (1+z)^{0.3}$, irrespective of the details of the stellar mass/dark matter profile or of the exact evolution in effective radius. The exception to this general trend is when a \cite{Burkert95} (rather than NFW) profile is used for the dark matter: in this case, a result similar to the no dark matter model (see \Fig\ref{fig:SigmaTracks} and following discussion) is obtained, due to the lower dark matter density in the central regions of the galaxy predicted by the \cite{Burkert95} profile. Our predicted evolution in velocity dispersion agrees well with the observational findings of \citet[][see also \citealt{Gargiulo15}]{Mason15}, who find $\sigmape \propto (1+z)^{0.20\pm0.07}$, and also consistent with \citet{Sande11} and \citet{Canna20}, who found $\sigmape \propto (1+z)^{0.4}$. Some previous semi-empirical models \citep[e.g.,][]{Hopkins09} predicted a very similar evolution in velocity dispersion at fixed stellar mass to our no-$\mstare/L$-gradient models. 

\begin{figure*}
	\includegraphics[width=\textwidth]{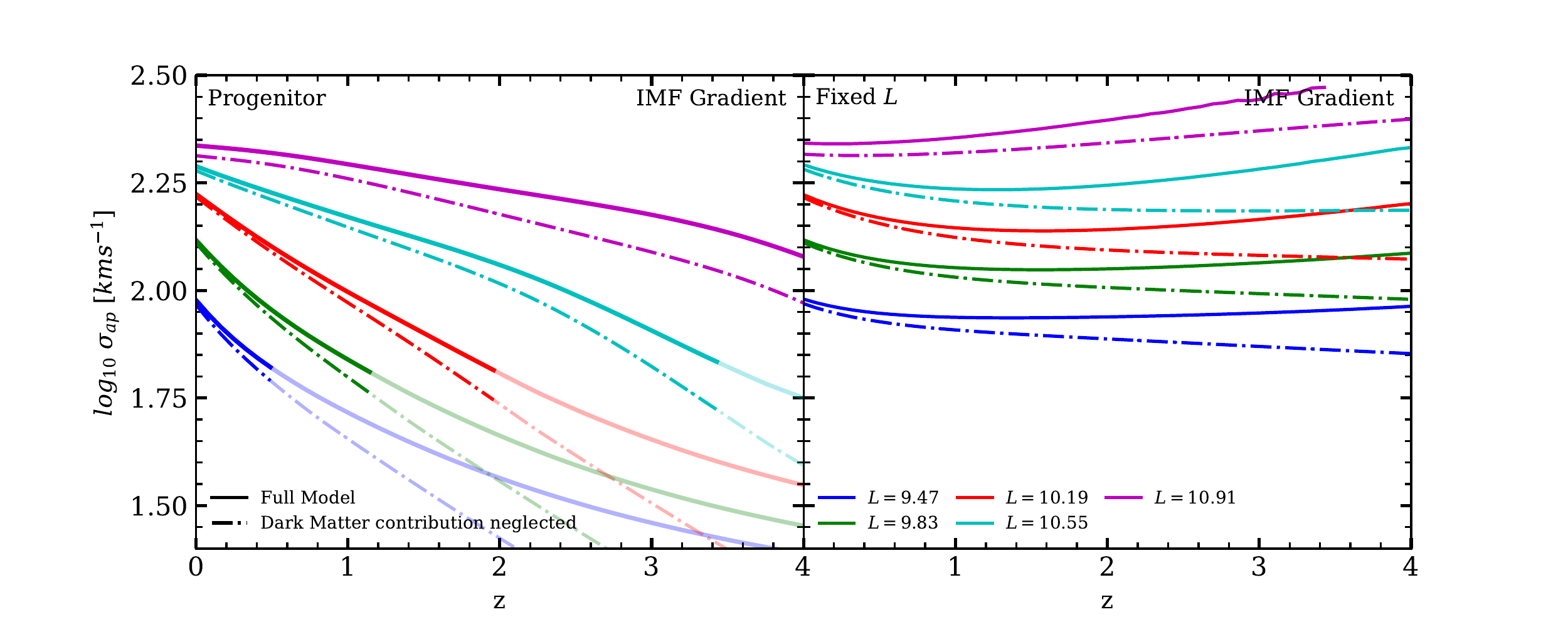}
    \caption{Average velocity dispersion histories for our fiducial model with an IMF driven $\mstare/L$ gradient. Colors correspond to the galaxies with total luminosities that would correspond to the usual bins of stellar mass (e.g. \Fig\ref{fig:SigmaTracks}) if the constant $\mstare/L$ model is assumed.}
    \label{fig:SigmaTracksIMF}
\end{figure*}
\begin{figure*}
	\includegraphics[width=\textwidth]{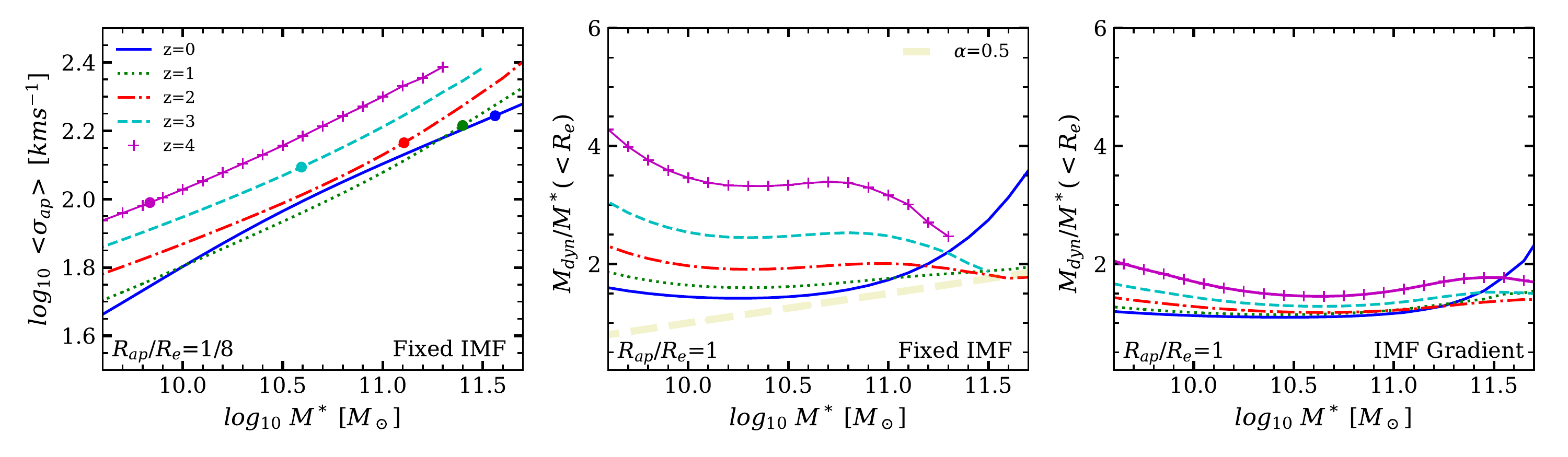}
    \caption{Predictions derived from our fiducial model of the evolution of the Faber-Jackson relation and the $M_{dyn}/\mstare$ ratio as a function of stellar mass and redshift for a fixed IMF and no $\mstare/L$ gradients (left and middle panels) and an IMF driven $\mstare/L$ gradient (right panel). In both models we find an increase of velocity dispersion at fixed stellar mass which is however less pronounced when there is an $\mstare/L$ gradient. In the left hand plot we also show the relative evolution of a galaxy of progenitor mass $log_{10}\,M^*\!=\!11.5M_\odot$, represented by the solid circles.}
    \label{fig:FJzTiltz}
\end{figure*}

To further highlight the significant role of dark matter in shaping the velocity dispersion evolutionary tracks, \Fig\ref{fig:SigmaTracks} compares the \sigmazM\ for different stellar masses along the progenitors (left panel) and at fixed stellar mass (right) with and without the contribution of dark matter in the mass budget (\Eq\ref{eq:massbudget}, solid and dotted lines, respectively). Of course, in this exercise we can only rely on our analytic semi-empirical models and cannot compare with the TNG50 hydrodynamic simulation, as in the latter it is not possible to remove the dark matter contribution from the gravitational potential and from the stellar particles' velocities (\Eq\ref{eq:sigmaTNG}), once more proving the flexibility and usefulness of semi-empirical models as exploratory tools. Velocity dispersions are predicted to rapidly drop along the main progenitors in a nearly parallel fashion when dark matter is removed (dot-dashed lines, left panel), at least for galaxies with $\log_{10} \mstare/M_\odot \lesssim 10^{11}$. More massive galaxies tend to retain their weak evolution with redshift as the stellar component dominates the inner dynamics for longer periods of time before it crosses the knee of the SMHM relation. Removing the dark matter component also flattens out the \sigmazM\ at fixed stellar mass (dot-dashed lines, right panel). As expected, when dark matter is not considered in the mass budget (\Eq\ref{eq:massbudget}), a steady decrease of stellar mass naturally implies a constant decrease in the corresponding velocity dispersion. Equivalently, at fixed stellar mass the model predicts an approximately constant velocity dispersion, further proving the relatively weak roles of effective radius and \Sersic\ index in determining central velocity dispersions. We note that we have neglected the contribution of gas in the computation of velocity dispersions throughout (see \Sec\ref{Method}), also given the sparsity of available data. Nevertheless, gas fractions in galaxies are predicted and observed to increase at higher redshifts \citep[e.g.,][and references thereafter]{Stewart09}, thus possibly promoting an even weaker evolution of velocity dispersions along the progenitors and an even steeper evolution at fixed stellar mass than the trends reported in \Fig\ref{fig:SigmaTracks}.

\Fig\ref{fig:SigmaTracksIMF} shows the analogue \sigmazM\ evolutionary tracks along the progenitors (left) and at fixed stellar mass (right) for models with an IMF driven $\mstare/L$ gradient, for different bins of galaxy luminosity $L$, as labelled. Similarly to when $\mstare/L$ gradients are ignored, the \sigmazM\ continue to present a marked downsizing, but their evolution is similar with (solid lines) and without (dot-dashed lines) dark matter at least up to $z\sim 1-2$ before diverging from one another at higher redshifts. This behaviour is expected from our discussion of \Fig\ref{fig:FP}, which supports the fact that a significant gradient in $\mstare/L$ could account for the full dynamical mass within $R_e$. Therefore, in the case of a variable IMF, the stellar mass tends to dominate over dark matter for a longer time before the dark matter is able to take over the gravitational budget of the central regions of the galaxy. The evolution of \sigmazM\ at fixed stellar mass is also similar to the case with no $\mstare/L$ gradients, but is somewhat weaker, roughly described by the scaling \sigmazM$\propto(1+z)^{0.2}$, which is, interestingly, in even better agreement with the observational results by \citet[][]{Mason15}.  

The behaviour of \sigmazM\ with redshift summarised in \Fig\ref{fig:SigmaTracks} helps to make solid predictions on other relevant galaxy probes which we introduced in \Sec\ref{Results-Low}. The left panel of \Fig\ref{fig:FJzTiltz} shows the predicted FJ relation at different redshifts. As velocity dispersion increases at fixed stellar mass, the FJ will in turn increase in normalization, consequent to the increasing central dark matter mass density. More specifically, the evolution in normalization is relatively weak up to $z\sim 1$, in agreement with some observational data \citep[e.g.,][]{Zahid16}, and then it starts increasing at a rate of $\propto (1+z)^{0.3}$, faithfully mirroring the evolution in \sigmazM\ at fixed stellar mass shown, e.g., in the right panel of \Fig\ref{fig:SigmaTracks}.
We confirm that we see a similar redshift evolution in the FJ relation extracted from the TNG simulation. In \Fig\ref{fig:FJzTiltz} we also include the locations of a galaxy with progenitor mass $log_{10}\,M^*\!=\!11.5M_\odot$, which nicely shows how a typical massive galaxy, whilst growing in stellar mass and velocity dispersion, moves ``up'' the FJ as redshift decreases. The middle panel of \Fig\ref{fig:FJzTiltz} reports the predicted evolution of the ratio between dynamical mass and stellar mass within the effective radius $R_e$, i.e. $M_{dyn}(<R_e)\propto \mstare(R_e)^{1+\alpha}$, in the same format as in \Fig\ref{fig:FP}. It is first of all evident that the normalization of the $M_{dyn}(<R_e)/\mstare(R_e)$ ratio increases with redshift consequent to the increase in the inner dark matter density which drives the increase in velocity dispersion. More interestingly, the initial tilt of $\alpha\sim 0.5$ at $z=0$ (as shown in \Fig\ref{fig:FP}) rapidly drops, and in fact $\alpha$ becomes negative at higher stellar masses and at high redshift. The main reason behind this strong and rapid evolution in this ratio has to be mostly ascribed, in our models, to the steady decrease in effective radii with redshift which is more pronounced in more massive galaxies, as parameterised in \Eqs\ref{eq:rez} and \ref{eq:gammaz} (see \Figs\ref{fig:FJz0} and \ref{fig:Gamma}). As the galaxy effective radii gradually shrink at earlier epochs, the contribution of dark matter becomes proportionally less relevant in massive galaxies. A similar trend in the shape of the $M_{dyn}(<R_e)/\mstare(R_e)$ ratio with time is predicted also in the presence of a variable IMF (right panel), though the normalization has a much weaker evolution in this instance, a trend once more mainly induced by the dominance of the stellar mass within $R_e$. 

\section{Discussion}
\label{sec:Discussion}

By using a semi-empirical approach, following dark matter assembly histories and making use of an input SMHM relation, we have built a flexible machinery to predict the velocity dispersion evolutionary tracks \sigmazM\ of galaxies in the stellar mass range $9<\log \mstare/M_\odot<11.5$. We showed that, in models with no $\mstare/L$ gradient, which appear very similar to the predictions of the TNG50 hydrodynamic simulation (\Fig\ref{fig:SigmaTracks}), the inner relative fraction of dark matter with respect to stars regulates the \sigmazM\ tracks, although its role becomes less prominent in the presence of sufficiently steep (IMF driven) gradients in $\mstare/L$. In this section we discuss our findings of \sigmazM\ evolutionary tracks that are constant or increasing with time, in light of galaxy evolutionary models, in particular focusing on role of ``dry'' (gas-poor) mergers, and on the link between velocity dispersion and central black hole mass. 

\subsection{The role of (dry) mergers}
\label{sec:DryMergers}

One of the key predictions from all hierarchical models of galaxy formation is that velocity dispersion should \emph{decrease} with cosmic time in the presence of minor, dry mergers (see \Sec\ref{sec:Intro}). A simple approximation to velocity dispersion evolution under dry mergers, derived from basic energy conservation arguments, can be written as \citep[e.g.,][]{Naab09,Fan10,Nipoti12,Lapi18}
\begin{equation}
\sigma_{after}^2=\sigma_{before}^2\frac{1+\eta^{2-\delta}}{1+\eta}\, ,
    \label{eq:sigmaLapi}
\end{equation}
where $\sigma_{before}$ and $\sigma_{after}$ are, respectively, the velocity dispersions of the central galaxy before and after a merger, $\eta$ is the stellar mass ratio between the infalling and central galaxy, and $\delta$ is the exponent of the radius-mass relation $r\propto M^{\delta}$ (which we set to a reference value of $\delta=0.6$ following \citealt{Shankar2014}).
To simulate the cumulative impact of mergers on velocity dispersions, we apply \Eq\ref{eq:sigmaLapi} to the mergers occurring along the dark matter main progenitor branches of the stellar mass accretion tracks that we generated as the baseline for our \sigmazM\ evolutionary tracks (\Fig\ref{fig:SigmaTracks}). More specifically, we follow the methodology put forward by \citet{Grylls2019STEEL,Grylls20SFR,Grylls20mergers}, and further developed and refined by Fu et al. (in prep.), in which infalling dark matter subhaloes at each interval $dz$ are extracted from the unevolved (i.e., unstripped), subhalo mass function \citep[e.g.,][]{Hopkins09,Jiang2016}. Galaxy stellar masses are assigned to both central and satellite dark matter haloes via a SMHM relation, and are then assumed to be ``frozen'' after infall, i.e. with negligible stripping and star formation once they cross the virial radius, which \citet{Grylls2019STEEL} showed to be a very good approximation for reproducing the local satellite stellar mass function above $\mstare \gtrsim 10^{10}\, M_\odot$, the range of mass of interest here. We stress that none of the conclusions discussed below will be qualitatively altered by allowing for some late evolution in the infalling satellites or by varying the rate of galaxy mergers via, e.g., the adoption a different input SMHM relation \citep[e.g.,][]{Grylls20mergers}. Our aim here is to check whether a sequence of repeated dry mergers, at the rate predicted by a $\Lambda\!CDM$ Universe, induces a (late) $\sigma$ evolution roughly consistent with the one inferred by our semi-empirical models.  

\begin{figure}
	\includegraphics[width=\columnwidth]{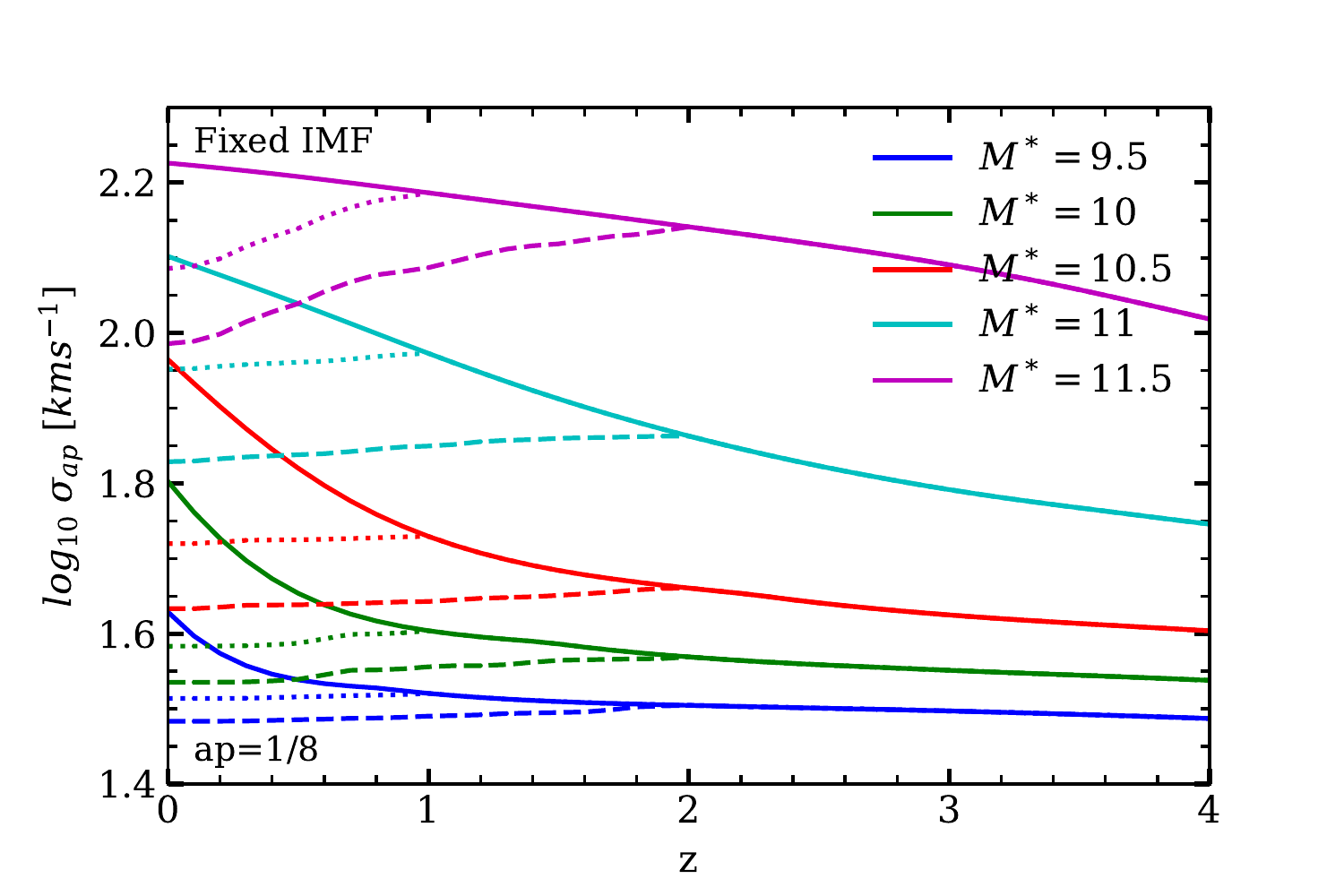}
    \caption{Average $\sigma$ for galaxies of fixed mass at $z=0$, showing the effect of multiple repeated mergers on the evolution of $\sigma$ commencing at $z=2$ and $z=1$. Velocity dispersion drops unreasonably quickly, suggesting other processes are at work.}
    \label{fig:SigmaLapi}
\end{figure}
\begin{figure*}
	\includegraphics[width=\textwidth]{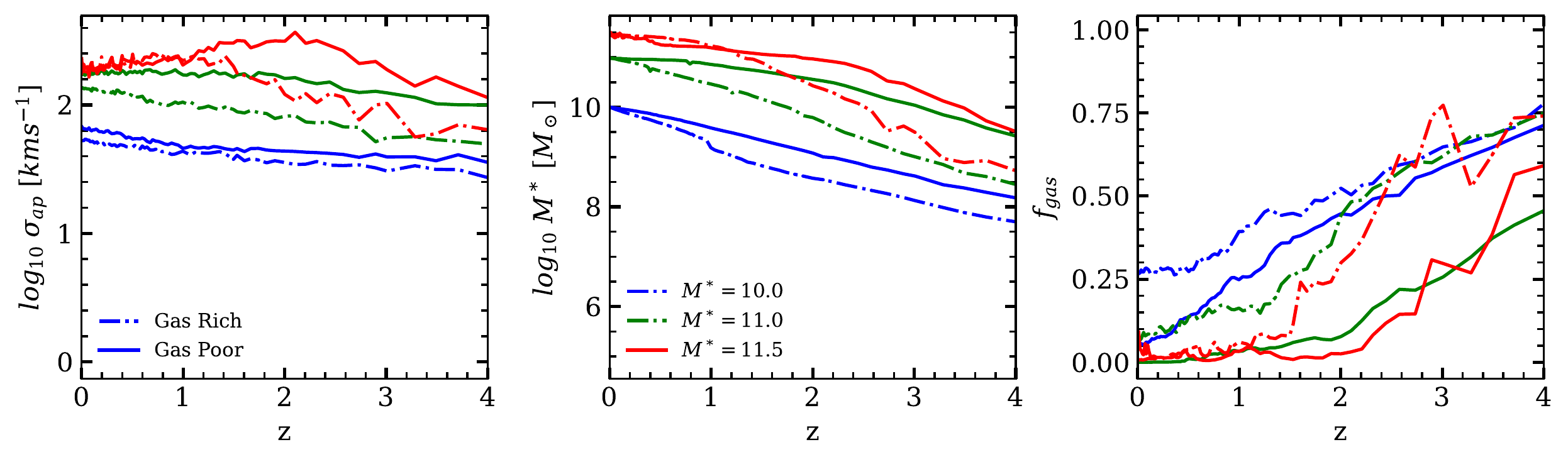}
    \caption{Velocity Dispersion, $\mstare$ and $f_{gas}$ histories from galaxies defined as Gas Rich and Gas Poor within TNG50. Galaxies are selected in mass bins of 0.1 dex (at $z=0$) and for each the ratio of gas to stellar mass within the half mass radius is computed. The upper and lower quartiles of this distribution are plotted here, as `Gas Rich' and `Gas Poor' respectively.}
    \label{fig:TNGGas}
\end{figure*}

The solid lines in \Fig\ref{fig:SigmaLapi} correspond to our \sigmazM\ evolutionary tracks in our fiducial model with a constant $\mstare/L$ for different stellar masses at $z=0$, as labelled. We assume that the galaxy undergoes a two-phase evolution \citep[e.g.,][]{Oser10,Lapi18}, comprising of an early in-situ growth followed by a sequence of mostly dry mergers. We choose the latter phase to kick in either at $z=1$ or at $z=2$, and thus from that redshift onwards, for each of our evolutionary tracks, we update the velocity dispersion following each merger event via \Eq\ref{eq:sigmaLapi}. The result is shown with dotted and dashed lines, respectively. For galaxies with stellar mass $\mstare \lesssim 10^{11}\, M_\odot$, the evolution of \sigmazM\ is predicted to be approximately flat from $z=1-2$, as naturally expected from \Eq\ref{eq:sigmaLapi} as galaxy mergers are progressively less relevant at lower stellar masses \citep[e.g.,][]{DeLucia2006,Hopkins2010mergers,Shankar13sizes}. Such a flat trend is in tension with the increasing \sigmazM\ at $z\sim0.5-1$, suggesting that in-situ growth is a dominant component for the mass and dynamical evolution of lower mass galaxies. More massive galaxies with $\mstare \gtrsim 10^{11}\, M_\odot$ instead show relatively flat \sigmazM\ at $z\sim 1-2$, which however are still in tension with the evolution via \Eq\ref{eq:sigmaLapi} which predicts velocity dispersions steadily decreasing with cosmic time, especially for galaxies $\mstare \gtrsim 2\times 10^{11}\, M_\odot$. We note that adopting a gradient in $\mstare/L$ would induce even steeper evolution in the predicted \sigmazM\ at all epochs (\Fig\ref{fig:SigmaTracksIMF}), thus exacerbating the tensions with pure dry merger models. We thus conclude that completely dry mergers cannot dominate the late evolution of galaxies, and that additional in-situ stellar mass growth, even in the most massive galaxies, should occur during and/or in between mergers to maintain a flat or increasing \sigmazM. This conclusion is consistent with the declining, but still non-zero, star formation histories inferred in massive galaxies \citep[e.g.,][and references therein]{BuchanShankar,Leja19,Grylls20SFR} and with the presence of substantial amounts of gas in massive galaxies at intermediate epochs \citep[e.g.,][]{Gobat18,Gobat20}. 

We note that our current model provides only average trends of velocity dispersion as a function of stellar mass, but we do expect a variety of evolutionary trends for galaxies of similar stellar mass. In addition, our semi-empirical models do not explicitly include any gas component. Gas fractions are observed to generally increase at earlier epochs at fixed stellar mass \citep[e.g.,][and references therein]{Stewart09}, and have sometimes been included in semi-empirical models to deal with galaxy mergers \citep[e.g.,][]{Hopkins2010mergers,Zavala12,Shankar2014}. Irrespective of the exact density profile chosen for the gas component within the host galaxy, the contribution of an increasing gas fraction to the velocity dispersion will become progressively more relevant with increasing redshift, similarly to what occurs for the dark matter component, as discussed in \Sec\ref{Results-High}. Thus, the inclusion of a gas component will, if anything, strengthen the results put forward here and in \Sec\ref{Results-High}, by further flattening the \sigmazM tracks along the progenitors and steepening them at fixed stellar mass.

To further probe the role of scatter and gas richness in shaping the evolutionary histories of \sigmazM, we dissect the velocity dispersion evolutionary tracks in the TNG50 simulation for different gas fractions and stellar masses. The left panel of \Fig\ref{fig:TNGGas} plots the \sigmazM\ of galaxies of equal stellar mass at $z=0$ (middle panel), but distinct gas fractions at all relevant epochs (right panel), namely the 95\% percentile above and below the mean gas fractions for galaxies of the selected stellar mass bin (dot-dashed lines and solid lines, respectively). We find clear evidence that gas-richer galaxies, as expected, have a significantly steeper evolution in both stellar mass and velocity dispersion, whilst gas-poorer galaxies tend to have a flatter evolution in stellar mass, and a flat or even decreasing velocity dispersion at late epochs, consistently with what predicted from our toy models in \Fig\ref{fig:SigmaLapi} (dashed lines). Interestingly, the most massive and gas-poor galaxies in the TNG50 simulation tend to show a decreasing velocity dispersion at late times $z\lesssim 2$ (solid red line in the left panel of \Fig\ref{fig:TNGGas}), which is what it would be expected in the presence of repeated dry mergers. In order to reproduce this trend in \sigmazM\ in our semi-empirical model we would require an input SMHM relation that steepens at earlier epochs, in a way to keep the $\zeta=\mstare(R_e)/[\mstare(R_e)+M_{halo}(R_e)]$ ratio (\Fig\ref{fig:SMHM}) increasing at earlier epochs along the halo progenitor track, a trend that is however not favoured by current estimates of the stellar mass function at high redshifts \citep[e.g.,][]{kawi2020}. Our conclusions agree with previous analytic and numerical work suggesting that dry mergers alone cannot entirely account for the behaviour of velocity dispersion at late epochs \citep[e.g.,][]{Nipoti12}. 

\subsection{Implications for black hole scaling relations}

In this Section we attempt to derive the $M_{bh}\!-\!\sigma$ at different redshifts following a methodology similar to the one put forward by other groups in the context of the $M_{bh}\!-\!\mstare$ relation \citep[e.g.,][]{Yang19,Carraro20,Shankar20}. We calculate the black hole accretion rate $\dot{M}_{bh}[z,\mstare]$ integrating at each time step the probability $P(z,L_X|\mstare)$ of having a certain X-ray AGN luminosity for a given stellar mass and converting to an accretion rate via the equation,
\begin{equation}
\begin{split}
\dot{M}_{BH}(\mstare, z) = \\
 \int_{-2}^\infty P(L_{SX}|\mstare, z) \frac{(1\!-\!\epsilon\!-\!\epsilon_{kin})k_{bol}(\mstare L_{SX})\mstare L_{SX}}{\epsilon c^2}dlog L_{SX}
\end{split}
\label{eq:Mbhz}
\end{equation}
where $P(L_{SX}|\mstare, z)$ is the probability distribution of specific X-ray luminosity, $\epsilon$ is the mean radiative efficiency (we set this to a nominal value of 0.1), $\epsilon_{kin}$ is the kinetic efficiency (we set this to 0.15, see, e.g., \citealt{Shankar09} and references therein), and $k_{bol}$ is the bolometric correction. In \Eq\ref{eq:Mbhz}, the (specific) X-ray luminosity is defined in units of host stellar mass, hence the lower limit of integration, -2, corresponds to $L_X\sim2\times10^{41}\, {\rm erg s^{-1}}$. For the functions $P$ and $k_{bol}$, we use the definitions described in \cite{Yang19}.
\begin{figure}
	\includegraphics[width=\columnwidth]{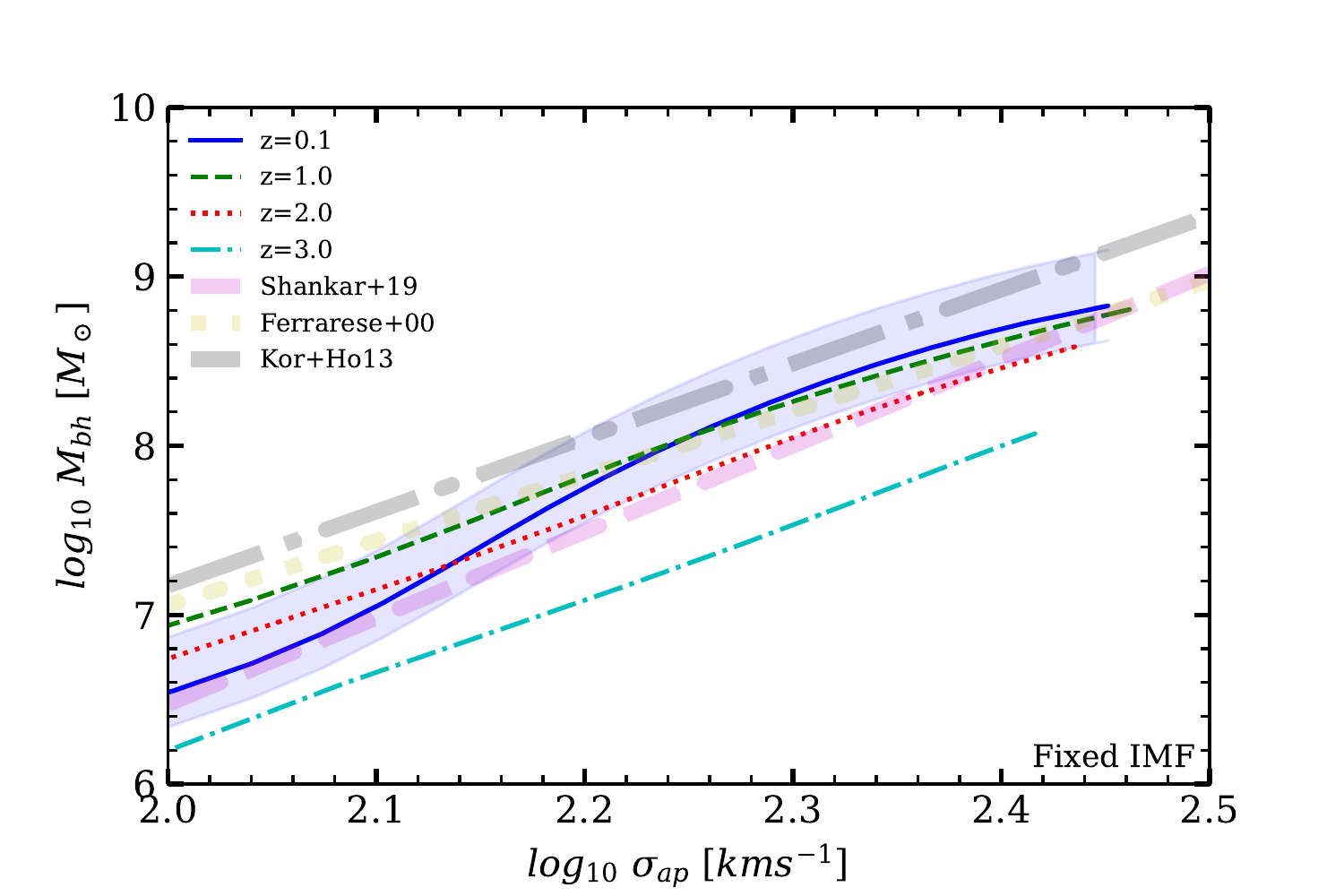}
    \caption{The $M_{bh}\!-\!\sigma$ relation as predicted by our model at different redshifts and compared with various relations at $z=0$ from the literature. The blue shaded region shows the effect of varying the radiative efficiency $\epsilon$ (at $z=0$) by $\pm0.05$.}
    \label{fig:BlackHoles}
\end{figure}
We then integrate in time the black hole accretion rate derived from \Eq\ref{eq:Mbhz} along each \sigmazM\ evolutionary track, which corresponds to a given stellar mass growth history $\mstare[z]$, to generate a corresponding $M_{bh}[z,\mstare]$ and thus ultimately building a $M_{bh}[z]\!-\!\sigma[z]$ relation. 

Our result is shown in \Fig\ref{fig:BlackHoles} for different redshifts, as labelled. We find a $M_{bh}\!-\!\sigma_{ap}$ relation that is roughly constant in both slope (which is around a value of $5$) and normalization up to $z\sim 2$, and possibly decreasing in normalization only beyond this epoch, in good agreement with available observations up to $z\sim 1-2$ \citep[e.g.,][]{Shankar09,SalvianderShields,Shen15}. When adopting reference values of $\epsilon=0.1$ and $\epsilon_{\rm kin}=0.15$ \citep[e.g.,][and references therein]{Shankar20}, our predicted $M_{bh}\!-\!\sigma$ relation at $z=0$ is in the ballpark of the one calibrated in the local Universe from the available sample of supermassive black holes with dynamical mass measurements \citep[e.g.,][]{Ferrarese2000,KormendyHo2013}. We note that the local data are less dispersed in the $M_{bh}\!-\!\sigma$ plane than in the $M_{bh}\!-\!\mstare$ one \citep[e.g.,][]{Shankar20Nat}, thus pointing to the former as a stronger constraint for galaxy evolution models \citep[e.g.,][]{Lapi18,Li20}. The modelling of the black hole accretion tracks $M_{bh}[z]$ via \Eq\ref{eq:Mbhz} generates a downsizing in black hole growth, as already found by \citet{Shankar20}, which mimics the one found here in velocity dispersion, further suggesting a close link between galaxies and their central black holes.  

In more general terms, the black hole accretion rate could be written as \citep[e.g.,][]{AllerRichstone,Marconi04,Shankar04}
\begin{equation}
    \label{eq:growth}
    \left\langle\frac{dM_{bh}}{dz}\right\rangle(M_{bh}, z) = \frac{U(M_{bh}, z)\lambda(z,Mbh)M_{bh}(z)}{t_{ef}}\frac{dt}{dz}dz
\end{equation}
in terms of the duty cycle $U(M_{bh}, z)$, i.e. the associated probability of a black hole of being active, and an Eddington ratio distribution $\lambda(M_{bh}, z)$ which, together with the e-folding time
\begin{equation}
    t_{ef} = \frac{\epsilon}{(1-\epsilon)}\,4\times10^8\, yr\, ,
\end{equation}
controls the average growth rate of a black hole at a given epoch. \Eq\ref{eq:growth} clearly highlights the degeneracy between duty cycle and Eddington rate in shaping black hole growth \citep[e.g.,][]{Shankar13acc,Aversa2015,Allevato21}. 
If we adopt an average duty cycle $U(M_{bh},z)$ as suggested by \citet[][their \Fig7]{Aversa2015}, in turn derived by continuity equation arguments and physically motivated black hole light curves, we find that in order for the black holes to remain on a non-evolving $M_{bh}\!-\!\sigma$, they must steadily reduce their Eddington ratios with cosmic time, in line with a number of independent measurements and observations \citep[e.g.,][and references therein]{Shankar13acc}.

\section{Conclusions}\label{sec:Conclu}

In this work we have presented a comprehensive analytic Jeans modelling to probe the galaxy scaling relations involving velocity dispersion \sigmap\ in the local Universe, as well as making valuable predictions for the evolutionary tracks \sigmazM\ along the progenitors and at fixed stellar mass. We compared our model predictions with a large sample of local MaNGA galaxies with spatially resolved velocity dispersions, and also with the outputs of the TNG50 hydrodynamic simulation. Our results can be summarised as follows:
\begin{itemize}
    \item When including both a bulge and a disk component in our models, we find that, in agreement with what measured in MaNGA, at fixed stellar mass velocity dispersions are relatively flat (\Fig\ref{fig:IndirectSim}) and become largely independent of the bulge-to-total ratio $B/T$ for $B/T\gtrsim 0.25$, irrespective of the exact stellar mass bin or aperture considered (\Fig\ref{fig:BTsig}), or whether or not a gradient in $\mstare/L$ is included.
    \item All our models can reproduce the full Faber-Jackson (FJ) relation, which is observed to be weakly dependent on galaxy morphology (\Fig\ref{fig:FJz0}). 
    \item The dynamical-to-stellar mass ratio $M_{dyn}/{\mstare}\sim {\mstare}^{\alpha}$ can be fully accounted for by an IMF driven gradient in the stellar $\mstare/L$ (\Fig\ref{fig:FP}).
    \item The predicted \sigmazM\ evolutionary tracks show, irrespective of the exact input parameters, a clear sign of downsizing, with more massive galaxies reaching their final \sigmap\ value at earlier epochs, whilst \sigmazM\ steeply increases approximately as $\sigma_{ap} \propto (1+z)^{0.3}$ at fixed stellar mass at constant $\mstare/L$. Very similar results are found when extracting velocity dispersions from the TNG50 simulation (\Fig\ref{fig:Size3x2}). 
    \item We interpret these behaviours in \sigmazM\ in light of the ratio between stellar mass to dark matter mass in the inner regions. When a galaxy falls below the knee of the stellar mass--halo mass relation, a decrease in halo mass implies a stronger decrease in stellar mass, thus the dark matter component becomes the dominant factor in controlling sigma (\Fig\ref{fig:SMHM}).
    \item The inclusion of a gradient in $\mstare/L$ maintains the downsizing in \sigmazM, but the single evolutionary tracks along the progenitors are steeper, whilst the ones at fixed stellar mass/galaxy luminosity evolve more slowly as $\sigma_{ap} \propto (1+z)^{0.2}$ or less (\Fig\ref{fig:SigmaTracksIMF}).
    \item The FJ relation is seen to increase in normalization, but not in slope, at the same pace as the dynamical-to-stellar mass ratio, closely following the degree of evolution of $\sigma_{ap}$ at fixed stellar mass  (\Fig\ref{fig:FJzTiltz}).
    \item Pure dry merger models are inconsistent with our \sigmazM\ evolutionary tracks (\Fig\ref{fig:SigmaLapi}), calling for additional processes, most probably residual star formation, as an important ingredient in shaping velocity dispersion through time. Indeed, we find that in the TNG50 simulation gas-rich galaxies at fixed stellar mass have a much steeper evolution in \sigmazM\ than their gas-poor counterparts of the same stellar mass (\Fig\ref{fig:TNGGas}). 
    \item For each \sigmazM\ we compute the growth of the central supermassive black hole derived from the $L_X-\mstare$ relation of stacked X-ray AGN. Our resulting $M_{bh}-\sigmape$ relation appears to be nearly independent of redshift, at least up to $z\lesssim 2$, with a constant slope of $\sim 5$ (\Fig\ref{fig:BlackHoles}).
\end{itemize}
All in all, our results point to a complex interplay between mergers and gas accretion in keeping velocity dispersion evolutionary tracks relatively flat or even increasing with cosmic time, a condition that supports the view in which velocity dispersions may retain memory of the initial stages of galaxy evolution. Our methodology proves the usefulness of data-driven semi-empirical models as \emph{complementary} tools to provide fast and robust predictions on galaxy properties and to probe the interplay of the different components (e.g., dark matter versus baryonic matter), something that would be difficult to disentangle in, e.g., hydrodynamic simulations. 

\section*{Acknowledgements}

C. Marsden acknowledges the ESPRC funding for his PhD.  F. Shankar acknowledges partial support from a Leverhulme Trust Research Fellowship. 
M. Bernardi acknowledges partial support from NSF grant AST-1816330.  H. Fu acknowledges support from the European Union's Horizon 2020 research and innovation programme under the Marie Sk\l odowska-Curie grant agreement No. 860744. We thank the anonymous referee for a careful reading of the manuscript and for several suggestions that helped improving the presentation of our results. Special thanks also to Michele Cappellari for several conversations on IMF variations, to Gary Mamon for useful insights on the analytic modelling of the disk contribution to velocity dispersion, to Dylan Nelson for a number of clarifications on the outputs of the TNG50 simulation, and to Carlo Nipoti for useful discussions. The IllustrisTNG simulations were undertaken with compute time awarded by the Gauss Centre for Supercomputing (GCS) under GCS Large-Scale Projects GCS-ILLU and GCS-DWAR on the GCS share of the supercomputer Hazel Hen at the High Performance Computing Center Stuttgart (HLRS), as well as on the machines of the Max Planck Computing and Data Facility (MPCDF) in Garching, Germany.

\section*{Data Availability}
Data will be shared upon request to the authors.



\bibliographystyle{mnras}
\bibliography{example} 




\appendix

\section{Virial coefficients}
\label{AppA}

In this paper we have presented a comprehensive methodology for computing velocity dispersion. Our source code is publicly available at \url{github.com/ChrisMarsden833/VelocityDispersion}, along with the associated documentation. In addition, we here provide a convenient and comprehensive look-up Table of virial coefficients that can be used to compute the total dynamical mass within the effective radius for a constant $\mstare/L$. 

The velocity dispersion of a spheroid within an arbitrary aperture can be well represented by the following relation
\begin{equation}
    \frac{GM(<\!R_e)}{R_e} = \mathcal{F} \sigma_{ap}^2
\end{equation}
where $G$ is the gravitational constant, $R_e$ is the scale radius, $M(<\!R_e)$ is the total mass within $R_e$ and $\sigma_{ap}$ is the velocity dispersion within the aperture. While some approximations for $\mathcal{F}$ already exist \citep[e.g. ][]{Bernardi2018}, a more comprehensive approximation that also takes into account the roles of dark matter and velocity anisotropy $\beta$ can be expressed as \begin{equation}
    \mathcal{F} = \zeta\mathcal{K}\left(\frac{R_{ap}}{R_e}, n\right) + \zeta_{halo}\mathcal{L}\left(\frac{R_{ap}}{R_e}, n, c\right) +  \mathcal{N}\left(\frac{R_{ap}}{R_e}, n, \beta\right)
    \label{eq:F}
\end{equation}
where $\zeta=\mstare(<\!R_e)/M(<\!R_e)$ is the ratio of stellar mass to total mass within $R_e$ (see \Fig\ref{fig:SMHM}), and $\zeta_{halo} = M_{halo}(<\!R_e)/M(<\!R_e)$ is the ratio of dark matter mass to total mass within $R_e$, which is simply $\zeta_{halo} =1-\zeta$ when neglecting gas and black hole masses. The functions $\mathcal{K}$, $\mathcal{L}$ and $\mathcal{N}$ are represented by numerical approximations. We present some example values in Table~\ref{table:approx}, but we also include a code in the aforementioned repository to numerically generate these Tables over arbitrary domains at the desired resolution.

Each function depends on the ratio of the aperture size $ap$ to $R_e$. $\mathcal{K}$ additionally depends on the \Sersic\ index $n$ and $R_e$, $\mathcal{L}$ depends on $n$ and the halo concentration $c$, $\mathcal{N}$ depends on $\beta$, the anisotropy parameter. If dark matter and anisotropy are neglected, the functions $\mathcal{L}$ and $\mathcal{N}$ (respectively) can be set to zero. In this case \Eq\ref{eq:F} reduces to the form presented in \citet[][see their \Eq2]{Bernardi2018}. Note that there is a subtle difference between their \Eq\ and ours (when neglecting dark matter and anisotropy), as the left hand side of our \Eq\ requires the mass within $R_e$ rather than the total stellar mass. In Table~\ref{table:zetaapprox} we also include some useful tabulated values of $\zeta$ as a function of stellar mass and effective radius, which are strictly valid for stellar and dark matter mass profiles from, respectively, \citet{Prugniel997} and \citet{NFW1996}, and halo concentrations from \citet{Ishiyama2020}, as discussed in \Sec\ref{Method}.

\begin{table*}
    \label{IMF}
	\centering
	\caption{Table containing numerical values for approximation of equation \ref{eq:F}. $n$ is the \Sersic\! index, $c$ is the halo concentration parameter, and $\beta$ is the anisotropy parameter.}
    \begin{tabular}{c|c|c|c}
    \hline
    Aperture $/R_e$ & $\mathcal{K}$ & $\mathcal{L}$ & $\mathcal{N}$ \\
    
    \hline
    
    $1/8$ &
    $
    \begin{array}{c|c}
    n & \mathcal{K} \\
    \hline
    2 & 2.79 \\
    3 & 2.35 \\
    4 & 1.94 \\
    5 & 1.63 \\
    6 & 1.38 \\
    \end{array}  $ &
    
    $
    \begin{array}{cc|ccccccc}
      &  &  &  & c &  &  &  & \\
     &  & 5 & 6 & 7 & 8 & 9 & 10 & 11 \\
     \hline
     & 2 & 3.99 & 3.99 & 4.0 & 4.01 & 4.01 & 4.02 & 4.02\\
     & 3 & 4.0 & 4.03 & 4.05 & 4.07 & 4.1 & 4.12 & 4.14\\
    n & 4 & 3.58 & 3.61 & 3.64 & 3.67 & 3.7 & 3.73 & 3.76\\
     & 5 & 3.12 & 3.15 & 3.18 & 3.21 & 3.24 & 3.27 & 3.31\\
     & 6 & 2.72 & 2.74 & 2.77 & 2.8 & 2.83 & 2.86 & 2.89\\
    \end{array} $& 
    $
   \begin{array}{cc|cccccc}
     &  &  & & \beta &  &  \\
     &  & -0.15 & 0.0 & 0.1 & 0.25 & 0.4 \\
     \hline
     & 2 & 0.42 & 0.0 & -0.28 & -0.7 & -1.12\\
     & 3 & 0.25 & 0.0 & -0.17 & -0.45 & -0.74\\
    n & 4 & 0.15 & 0.0 & -0.11 & -0.28 & -0.48\\
     & 5 & 0.1 & 0.0 & -0.07 & -0.19 & -0.32\\
     & 6 & 0.07 & 0.0 & -0.05 & -0.13 & -0.22\\
    \end{array} $\\

     \hline
     
     1 & 
     $
     \begin{array}{c|c}
    n & \mathcal{K} \\
    \hline
    2 & 2.83 \\
    3 & 2.71 \\
    4 & 2.51 \\
    5 & 2.29 \\
    6 & 2.07 \\
    \end{array}$ &
     $
     \begin{array}{cc|ccccccc}
     &  &  &  &  c & &  &  & \\
     &  & 5 & 6 & 7 & 8 & 9 & 10 & 11 \\
     \hline
     & 2 & 2.53 & 2.55 & 2.57 & 2.59 & 2.6 & 2.61 & 2.63\\
     & 3 & 2.86 & 2.88 & 2.9 & 2.91 & 2.93 & 2.94 & 2.95\\
    n & 4 & 3.03 & 3.05 & 3.06 & 3.08 & 3.09 & 3.1 & 3.12\\
     & 5 & 3.06 & 3.07 & 3.09 & 3.11 & 3.12 & 3.14 & 3.15\\
     & 6 & 2.99 & 3.01 & 3.03 & 3.05 & 3.06 & 3.08 & 3.1\\
    \end{array}$ & 
     $
     \begin{array}{cc|ccccc}
     &  &  &  & \beta &  & \\
     &  & -0.15 & 0.0 & 0.1 & 0.25 & 0.4 \\
     \hline
     & 2 & 0.09 & 0.0 & -0.06 & -0.16 & -0.28\\
     & 3 & 0.07 & 0.0 & -0.05 & -0.14 & -0.23\\
    n & 4 & 0.06 & 0.0 & -0.04 & -0.11 & -0.19\\
     & 5 & 0.04 & 0.0 & -0.03 & -0.09 & -0.15\\
     & 6 & 0.04 & 0.0 & -0.03 & -0.07 & -0.12\\
    \end{array} $\\
     
     \hline
    \end{tabular}
    \label{table:approx}
\end{table*}

\begin{table*}
    \label{IMF}
	\centering
	\caption{Table containing numerical values of $\zeta$ based on our model, as a function of stellar mass and effective radius.}
    \begin{tabular}{cc|cccccccccccc}
\hline
&&&&&& $R_e\,[kpc]$ &&&&&&& \\
 & & 1.0  & 3.0  & 5.0  & 7.0  & 9.0  & 11.0  & 13.0  & 15.0  & 17.0  & 19.0  & 21.0  & 23.0 \\ 
\hline
 &  9.0  & 0.5 & 0.12 & 0.05 & 0.03 & 0.02 & 0.02 & 0.01 & 0.01 & 0.01 & 0.01 & 0.01 & 0.01 \\
 &  9.5  & 0.75 & 0.28 & 0.14 & 0.09 & 0.06 & 0.05 & 0.04 & 0.03 & 0.03 & 0.02 & 0.02 & 0.02 \\
 &  10.0  & 0.9 & 0.53 & 0.31 & 0.21 & 0.15 & 0.11 & 0.09 & 0.08 & 0.06 & 0.06 & 0.05 & 0.05 \\
$log_{10}$ $\mstare$ $ [M_\odot]$ &  10.5  & 0.96 & 0.75 & 0.55 & 0.4 & 0.31 & 0.24 & 0.2 & 0.16 & 0.14 & 0.12 & 0.11 & 0.1 \\
 &  11.0  & 0.98 & 0.88 & 0.73 & 0.6 & 0.48 & 0.4 & 0.33 & 0.28 & 0.24 & 0.21 & 0.18 & 0.16 \\
 &  11.5  & 0.99 & 0.94 & 0.85 & 0.74 & 0.64 & 0.55 & 0.47 & 0.41 & 0.35 & 0.31 & 0.27 & 0.24 \\
 &  12.0  & 1.0 & 0.97 & 0.92 & 0.86 & 0.78 & 0.71 & 0.64 & 0.57 & 0.51 & 0.46 & 0.41 & 0.37 \\
 \hline
    \end{tabular}
    \label{table:zetaapprox}
\end{table*}

\section{Stellar mass--halo mass relation} \label{appendix:hmsm}

We here briefly recall the parameterization of the stellar mass--halo mass relation from \citet{Grylls2019STEEL}, including the correction from \citet{Zanisi21}, which we use as a reference throughout the work. The analytic formula is 
\begin{equation}
\begin{array}{l}
\mstare(M_{halo}, z) = 2\frac{M_{halo} N(z)}{10^{0.1}} \left[\left( \frac{M_{halo}}{M_n(z)}\right)^{-\beta(z)} + \left( \frac{M_{halo}}{M_n(z)}\right)^{\gamma(z)} \right]^{-1}\\

\;\;N(z) = N_{0.1} + N_z \left(\frac{z - 0.1}{z + 1}\right)\\

\;\;M_n(z) = M_{n,\,0.1} + M_{n,\,z} \left(\frac{z - 0.1}{z + 1}\right)\\

\;\;\beta(z) = \beta_{0.1} + \beta_z \left(\frac{z - 0.1}{z + 1}\right)\\

\;\;\gamma(z) = \gamma_{0.1} + \gamma_z \left(\frac{z - 0.1}{z + 1}\right)\, .
\end{array}
\label{eq:GryllsHMSM}
\end{equation}
Here $\mstare$ represents the stellar mass of the galaxy and $M_{halo}$ represents the host halo mass. The values of the parameters $N$, $M_n$, $\beta$ and $\gamma$ are given in Table~\ref{tab:HMSMtable} (the subscript $0.1$ refers to the value of the parameters at $z=0.1$). The \citet{Grylls2019STEEL} relation is mostly valid in the redshift range $0.1<z<4$. 

\begin{table} 
	\centering
	\caption{Parameters for equation \ref{eq:GryllsHMSM}.}
	\label{tab:HMSMtable}
	\begin{tabular}{lccccr}
		\hline
		$ $ & $M_n$ & $N$ & $\beta$ & $\lambda$ & $\sigma$\\
		\hline
		Central, $z=0.1$ & 11.95 & 0.032 & 1.61 & 0.54 & 0.11\\
		Total, $z=0.1$  & 11.89 & 0.031 & 1.77 & 0.52 & 0.10\\
		\hline
		Evolution, $z > 0.1$ & 0.4 & -0.02 & -0.6 & -0.1 & N/A\\
		\hline
	\end{tabular}
\end{table}

\section{Determination of Stellar Velocity Dispersion from TNG}
\label{AppB}

As discussed in the main text, we utilize the results from the Illustris TNG simulation  \citep{TNGMain, TNGSup1, TNGSup2} to compare the evolution of the velocity dispersion. Here we briefly describe the process of extracting velocity dispersions from the TNG galaxies, as projected velocity dispersions within an aperture are not available in the provided datasets. Firstly we select the main progenitor history of all galaxies within an appropriate mass bin, and at each snapshot calculate $\sigma$ as follows. We first project the stellar particles belonging to the galaxy (determined using a FOF method, see the TNG documentation for more details) along an arbitrary direction. Next, to obtain realistic values for $R_e$ (and also $n$, although we do not utilize this in this paper), we fit a \Sersic\ profile to the projected density of these particles. Next, we eliminate all particles that are not within our projected aperture size. We then calculate the velocity dispersion of the remaining stellar particles in the projected dimension, weighted by their masses
\begin{equation}
    \sigma_{x} = \sqrt{\frac{\sum{m_i(v_{i, x}-\Bar{v}_{x})^2 }}{m_i}}\, ,
    \label{eq:sigmaTNG}
\end{equation}
where for particle $i$, $x$ is the projected direction  (hence $v_{i, x}$ is the component of the velocity in direction $x$). $\Bar{v}_x$ is the average (component of) velocity of all the particles, again weighed by the mass\footnote{Additional corrections must naturally be applied for the periodicity of the box, cosmology etc, as described in the TNG documentation.}. 
We compute the projected velocity dispersion for each galaxy three times, projecting in the directions $x, y$ and $z$ (in the simulation coordinate system), and take the mean of these values to minimize any bias due to the projection axis. 

\section{Extreme values of $\Upsilon\!(R)$}

We provide in Table~\ref{IMFgrad} the values of the scale-dependent mass-to-light ratio at $R_e$ ($\Upsilon_{0}$) and at the centre ($\Upsilon_{max}$) for galaxies of different absolute magnitudes and measured velocity dispersion. As described in \Sec\ref{Method}, we assume the $\mstare/L$ varies linearly between these two extreme values and it is constant, equal to $\Upsilon_{0}$, at radii $R>R_e$.  

\begin{table}
    \label{IMFgrad}
	\centering
	\caption{Table of the extreme values of the mass-to-light ratios values for galaxies of different absolute magnitudes and measured velocity dispersion.}
    \begin{tabular}{llll}
    \hline
    Absolute magnitude (r-band) & $log_{10}\,\sigma\, [kms^{-1}]$ & $\Upsilon_{max}$ & $\Upsilon_{0}$ \\
    \hline
    Ellipticals \\
    \hline
    
    \multirow{2}{*}{$-22.5 > M_r > -23.5$ $\begin{dcases*} \\ \end{dcases*}$}  & $2.4 - 2.5$  & 8.0 & 3.5 \\
    &  $2.3 - 2.4$  & 7.0 & 3.0 \\ 
    
     \multirow{2}{*}{$  -21.5 > M_r > -22.5 $ $\begin{dcases*} \\ \end{dcases*}$}  & $2.3 - 2.4$  & 5.0 & 3.0 \\
    &  $2.2 - 2.3$  & 5.0 & 2.5 \\ 
    
      \multirow{2}{*}{$  -20.5 > M_r > -21.5 $ $\begin{dcases*} \\ \end{dcases*}$}  & $2.2 - 2.3$  & 5.0 & 3.0 \\
    &  $2.1 - 2.2$  & 3.0 & 2.0 \\ 
    
    \hline
    S0s \\
    \hline
    
    \multirow{3}{*}{$-21.5 > M_r > -22.5$ $\begin{dcases*} \\ \\ \end{dcases*}$}  & $2.3 - 2.4$  & 6.5 & 4.5 \\
     &  $2.2 - 2.3$  & 5.0 & 2.0 \\ 
    &  $2.1 - 2.1$  & 2.0 & 1.0 \\ 
    
    \multirow{3}{*}{$-20.5 > M_r > -21.5$ $\begin{dcases*} \\ \\ \end{dcases*}$}  & $2.2 - 2.3$  & 5.5 & 2.0 \\
     &  $2.1 - 2.2$  & 3.0 & 1.5 \\ 
    &  $2.0 - 2.1$  & 2.0 & 1.5 \\ 
    
    \multirow{3}{*}{$-19.5 > M_r > -20.5$ $\begin{dcases*} \\ \\ \end{dcases*}$}  & $2.1 - 2.2$  & 4.0 & 3.0 \\
     &  $2.0 - 2.1$  & 4.0 & 3.0 \\ 
    &  $1.9 - 2.0$  & 1.5 & 1.5 \\

     \hline
    \end{tabular}
\end{table}

\section{The impact of segregation in the SSP phase-space distribution functions}

\begin{figure}
	\includegraphics[width=\columnwidth]{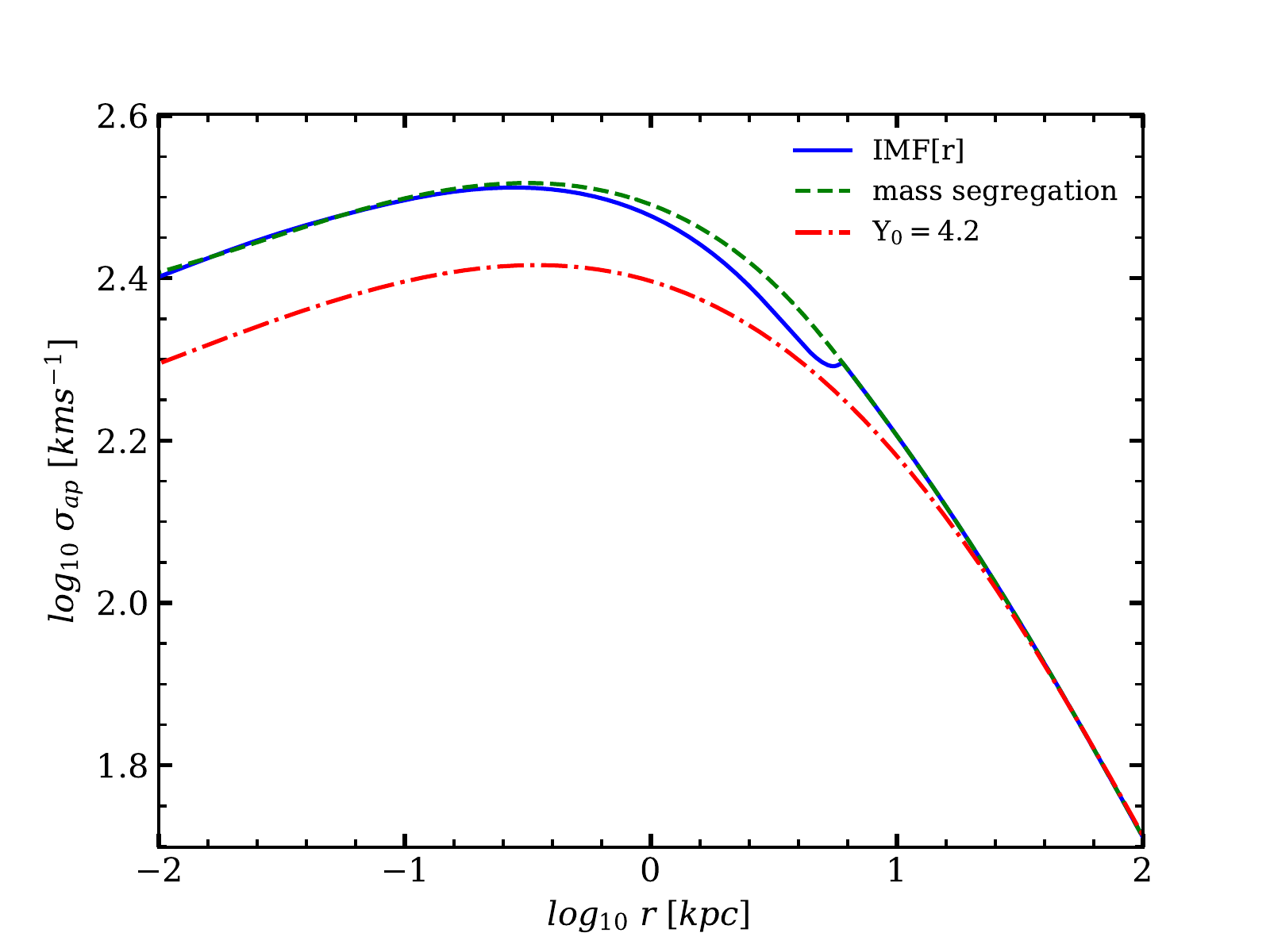}
    \caption{$\sigmale$ radial profiles for a typical galaxy in our MaNGA sample with $\phi=1.3$, $\Upsilon_0=3$, $\log L/L_\odot=11$, computed via \Eq\ref{eq:GeneralSolnAppStars} assuming both a constant $\mstare/L$ (red dot-dashed line) and a scale-dependent $\mstare/L$ as given in \Eq\ref{eq:ML} (blue solid line), and via \Eq\ref{eq:GeneralSolnAppStars2} with the same scale-dependent $\mstare/L$ (green dashed line).}
    \label{fig:AppSigma}
\end{figure}

The formulation in \citet{Bernardi2018}, which we followed in \Sec\ref{Method} to compute the radial profile and velocity dispersion, assumes that the objects (~low mass stars) which cause the IMF gradient are not dynamically different from the others (e.g., if stars always form in clusters, but the stellar IMF in the clusters depends on how far the cluster is from the centre of the galaxy). In this case, ignoring dark matter for the time being, \Eq\ref{eq:GeneralSoln} would read as 
\begin{equation}
    \label{eq:GeneralSolnAppStars}
    \rho^*(r)\sigma_r^2(r) = \frac{1}{f(r)}\int_{r}^{\infty} f(s) \rho^*(s) \frac{GM^*(s)}{s^2} ds
\end{equation}
with $\rho^*(r)=\rho^*_{Ser}(r) + \Delta \rho^*(r)$ and $M^*(r) = M^*_{Ser}(r) + \Delta M^*(r)$. Instead, if one thinks of a galaxy as being a linear combination of simple stellar populations (SSPs), each having its own $\mstare/L$ ratio, and that each SSP is described by its own \emph{distinct} phase-space distribution function, then one would rearrange the Jeans Equation in \Eq\ref{eq:GeneralSolnAppStars} in the following way
\begin{equation}
    \label{eq:GeneralSolnAppStars2}
    \rho^*_{Ser}(r)\sigma_r^2(r) = \frac{1}{f(r)}\int_{r}^{\infty} f(s) \rho^*_{Ser}(s) \frac{GM^*(s)}{s^2} ds\, ,
\end{equation}
where the low mass stars, similarly to a dark matter component, would not significantly contribute to the total luminosity but only to the total mass and thus would appear only on the right-hand side of \Eq\ref{eq:GeneralSolnAppStars2} \citep[see][]{Caravita21}. Our simple test shown in \Fig\ref{fig:AppSigma} for a given galaxy of with $\phi=1.3$, $\Upsilon_0=3$, $\log L/L_\odot=11$, indicates that, for $\mstare/L$ gradients of current interest, the two formulations only lead to relatively small differences in the $\sigmale$ profiles at $R\lesssim R_e$. We hope that future datasets will have sufficient signal-to-noise to determine which of the two formulations of the dynamics of the stars giving rise to IMF gradients is more realistic.

\bsp	
\label{lastpage}
\end{document}